\documentclass[english,12pt]{article}

\usepackage{epsfig}
\usepackage{cite}
\usepackage[T1]{fontenc}
\usepackage[latin1]{inputenc}
\usepackage{geometry}
\usepackage{graphicx}
\usepackage{amsmath,amssymb}

\makeatletter

\usepackage{amsmath,epsfig,bbm}

\usepackage{babel}
\makeatother
\begin{document}

\title{
\vspace{-2cm}
\begin{flushright}
\small{IPPP/03/78} \\
\small{DCPT/03/156}\\
\small{DESY 03-203}
\end{flushright}
\vspace{1cm}
\bf {Flavour in Intersecting Brane Models and Bounds on the String Scale}}
\author{\\
{\bf Steven A. Abel}\\ {\it IPPP, Centre for Particle Theory, University of Durham,}\\
{\it Durham DH1 3LE, UK}
\\ \\ 
{\bf Oleg Lebedev} \\ {\it DESY Theory Group, D-22603 Hamburg, Germany }
\\ \\ 
{\bf Jose Santiago}\\ {\it IPPP, Centre for Particle Theory, University of Durham,}\\
{\it Durham DH1 3LE, UK} }

\maketitle
\abstract{ We study flavour issues in nonsupersymmetric
intersecting brane models.
Specifically, the purpose of the present paper is twofold: (i) 
to determine whether realistic flavour structures can be obtained
in these models, and (ii) to establish  whether the non-supersymmetric
models address the gauge hierarchy problem.
To this end, we find that realistic flavour structures,
although absent at tree level, can 
arise even in the simplest 
models after effects of 4 fermion instanton--induced 
operators and radiative corrections
have been taken into account. On the other hand, our analysis
of  flavour changing neutral currents (FCNC), electric dipole moments (EDM),
supernova SN1987A  and other constraints shows that the string scale has to
be rather high, $10^4$ TeV. This implies that non-supersymmetric
intersecting brane models face a severe finetuning problem.
Finally, we comment on how non--trivial flavour structures can arise
in supersymmetric models.   }

\section{Introduction}

Models with D-branes intersecting at angles~\cite{Berkooz:1996km} 
have received a great deal
of attention due to their attractive phenomenological 
properties~\cite{Blumenhagen:2000wh}-\cite{Kors:2003wf}. In
particular, they have the potential to provide 
a nice geometric explanation  of the fermion family replication  (repeated
generations correspond to multiple D-branes intersections), 
the  Yukawa coupling hierarchy
(Yukawa couplings depend exponentially on the area 
spanned by the brane intersections), and so on.

Since supersymmetric models arise only for specific values of
intersection angles, realistic examples are relatively hard to come by, 
and many of the models that have been proposed are non-supersymmetric.
Despite outstanding theoretical issues such as stability, non-supersymmetric 
configurations are interesting
since they are a stringy realization of the
ADD idea \cite{Arkani-Hamed:1998rs}, with the string scale lying at just a few TeV.
This requires that some of the compactified dimensions be large,
which can be achieved in some brane constructions 
by, for instance, gluing a large volume manifold
accessible to gravity only~\cite{Blumenhagen:2002wn}.
Thus, it seems possible that the gauge hierarchy problem
can be addressed in non-supersymmetric models.

One known shortcoming of intersecting brane models 
comes from flavour physics,
that is, the Yukawa matrices are predicted to be
factorizable if the low energy theory is SM-- or MSSM--like,
e.g.
\begin{equation}
Y_{ij}=a_i b_j\;.
\label{factY}
\end{equation}
This is rank one and 
consequently only the third generation acquires mass.
The result is not affected by field theory 
renormalization group running.
One may therefore be tempted to dismiss such models 
at least as a way of generating a realistic fermion spectrum.

One of the purposes of this paper is to point out that there 
is always an additional
source of flavour structures in non-supersymmetric 
intersecting brane models.
This is the 4 fermion operators induced by string instantons.
Their flavour structure is not the same as that of the 
Yukawa interactions in the sense that they cannot be
diagonalized simultaneously. Through 1 loop threshold corrections
(which are independent of the string scale),
the 4 fermion operators contribute to the Yukawa couplings and
destroy the factorizability of the Yukawa matrices.
As a result, a realistic picture of the quark masses and mixings
can emerge. It is important to emphasize that this 
mechanism applies to the {\it non-supersymmetric} models only
due to SUSY non-renormalization theorems. For the supersymmetric
case, we point out another plausible mechanism based on
SUSY vertex corrections whose viability requires further study.

Another purpose of this paper is to determine 
whether 
intersecting brane models allow for a TeV string scale,
for which we utilize the realistic flavour structures alluded to above.
We employ the salient features of intersecting brane models,
such as the mechanism for family replication, the presence of
extra gauge bosons and so on, to obtain constraints on the string scale.
The strongest bounds stem from the FCNC constraints which
require the string scale to be no lower than $10^3 - 10^4$ TeV.
Other considerations such as EDMs, supernova cooling, etc.
allow for a lower string scale, ${\cal O}$(10) TeV.
Taken together, our results indicate that non-supersymmetric 
intersecting brane models suffer from a serious finetuning problem
and the ADD idea cannot be realized. Supersymmetry or some 
other solution to fine-tuning is still required for realistic models. 
In particular, we note that the bound of $10^3$ TeV applies for a 
whole host of independent flavour changing processes and  
it would be impossible to circumvent all of these bound by adjusting
relevant parameters.

The issue of flavour physics constraints on the string scale was
studied by two of us (S.A. and J.S.) in a previous
publication \cite{Abel:2003fk}. 
That analysis could be (and indeed was) critisized in that 
there was not at the time  a realistic 
model of flavour (due to the Yukawa factorization property)
that could be used as a starting point.
One might have argued  that 
whatever mechanism eventually generates flavour also aligns contributions 
to the FCNC processes in such a way that they are suppressed.
One of the points of the present work is to investigate
whether such an alignment could occur. Here we find no evidence for it, 
and, on the contrary, establish that  the bounds become  
significantly stronger than those presented in Ref.\cite{Abel:2003fk}.

The fact that the FCNC constraints become stronger (compared to those of 
Ref.\cite{Abel:2003fk}) once the  flavour model is specified
is natural and can be explained as follows.
The lower bound on the string scale in Ref.\cite{Abel:2003fk} was
derived by varying the string scale as a function of the
compactification scale.  This was an {\em absolute} bound
therefore: if we optimize the compactification scale to reduce FCNC, 
what is the minimum string scale that we can get away
with? 
However, the compactification scales at the optimized compactification
scale are typically not very realistic for gauge or
Yukawa couplings because they are rather large, and 
large compactification scales tend to dilute
the couplings with either large world volumes or large instanton
actions. 
Hence, with a full model of flavour there is no longer any
freedom to adjust compactification scales to reduce FCNC. 
The present work can be regarded as deriving 
the {\em typical} constraints
in a realistic flavour model which,
according to the argument above,
 should be stronger than the ones we obtained earlier.

In the following section, we begin by discussing 
contact interactions in intersecting brane models, and show that 
threshold effects can lead to a reasonable model of flavour. This 
1--loop calculation
 will be done in a field theoretical manner, by using the 
flavour changing tree-level 4 point interaction (derived in string theory) 
as an effective vertex. In principle, a full string calculation of the 
threshold effects is possible as well \cite{meben} but this 
will be left to future work. Following this, we will 
derive the FCNC constraints and find that they are enhanced over 
the absolute bounds in Ref.\cite{Abel:2003fk}. We then turn to non--FCNC 
constraints, i.e. 
EDM, astrophysical, LEP consraints on contact interactions and 
constraints on the $\rho$ parameter.  
In all cases, they are $subdominant$, 
and the FCNC constraints (in particular, those  from
the Kaon system, but also the $B$ and $D$ mesons) are paramount. 
Nevertheless, they 
provide additional support for the statement 
that TeV--scale intersecting brane models are highly unlikely. 
It  may be possible to fine-tune away one or a few  undesirable 
effects, 
but it is surely impossible to fine-tune 
away all of them. 

\section{Realistic flavour structures from contact terms}

\begin{figure}
\begin{center}\includegraphics[%
  scale=1.]{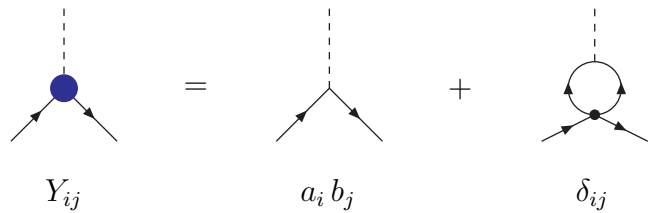}\end{center}
\caption{Threshold correction to Yukawa couplings. The black dot
in the loop diagram  
represents a chirality changing four fermion amplitude.
\label{cap:Threshold-correction-to}}
\end{figure}

In this section, we shall collect the components 
we need from the string theory calculations
of amplitudes. In particular, 
as well as the tree-level Yukawa coupling interactions, we will need 
flavour structures to be introduced via 
  four point couplings which are always present in this class of models. 
As we shall see, this naturally  circumvents the ``trivial
Yukawa'' (Eq.\ref{factY})  problem and provides a working model of flavour. 
The relevant diagram, shown in figure \ref{cap:Threshold-correction-to}, 
generates a threshold correction to the Yukawa coupling structure.
The blob in the loop diagram  is the tree-level four fermion 
flavour changing coupling, coming with a factor 
of $1/M_S^2 $,
which can be calculated in  string theory. 
The dimensionless Yukawa couplings 
get a significant (but loop--suppressed) contribution 
from this threshold correction if the effective cut-off
in the loop momentum is similar to $M_S$, as is natural. 
In supersymmetric theories, there are two contributions, 
with fermions and bosons in the loop which
cancel each other.
In non--supersymmetric theories, this is not the case.
For our purposes, it is sufficient to estimate this effect
in field theory using a hard cut-off although a string theory
calculation is also possible. (We shall discuss the procedure for 
performing the string calculation later.) What is 
important for us is that
the threshold correction has a ``nonfactorizable'' form,
\begin{equation}
Y_{ij}=a_i b_j + \delta_{ij}\;.
\end{equation}
Then, although the effect is loop--suppressed, it ``redistributes''
the large Yukawa coupling of the third generation and generates
relatively small masses (as well as mixings) for the lighter generations.

The desired contact interactions also contribute 
to the flavour changing processes. 
However, their contributions are  suppressed by a factor of $1/M_S^2 $. 
Consequently, the FCNC effects
decouple (unlike the threshold corrections) 
as the string scale is raised, and these interactions place 
a constraint of the string scale itself, as already seen in
Ref.\cite{Abel:2003fk}. 

The contact interaction that will be of major importance is 
the four fermion interaction $(\overline{q}q)(\overline{q}q)$
which can in field theory be  induced by a
Higgs boson exchange.  In string theory, these interactions 
can be much enhanced 
over the usual field theory amplitude suppressed by  two Yukawa couplings.
The key point can be summarized as follows. It is often assumed
that, once the Yukawas have been determined from the non-perturbative
\char`\"{}instanton\char`\"{} contribution, all other interactions
can be understood perturbatively at the level of effective field
theory. This assumption is incorrect if the string scale is low. 
In particular, for
contact interactions generated by a Higgs boson exchange, in
field theory, the amplitude is naturally proportional to
the product of two Yukawa couplings. 
In string theory, this is not necessarily the case.
As we shall see, the $s$- or $t$-channel 
Higgs exchange is extracted from the four fermion amplitude 
as a \char`\"{}double instanton\char`\"{} contribution 
(i.e. one instanton for each Yukawa coupling). 
However, the same 4 fermion operator may be generated by a 
single ``irreducible'' instanton, without going through
the Higgs vertex. The corresponding amplitude will dominate
if the single instanton action is significantly smaller than the 
double instanton action, i.e. if the area swept by propagating  open
strings is minimal for the former. 
This effect can be used to 
place strong constraints on $M_S$ from observables  
such as electric dipole moments which are normally Yukawa--suppressed
and plays a significant role in our discussion. 

\begin{figure}
\begin{center}\includegraphics{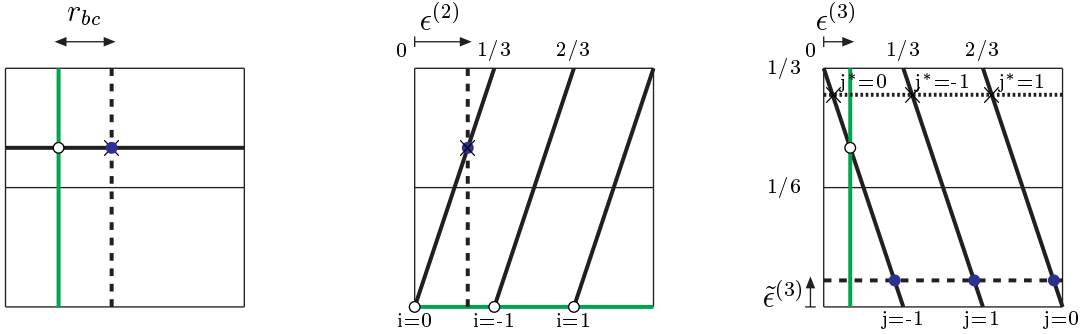}\end{center}
\caption{Brane configuration in a model of D6-branes intersecting at angles.
The leptonic sector is not presented here while the baryonic, left, right 
branes and
orientifold image of the right brane 
are  the dark solid, faint solid,
dashed and dotted lines, respectively. 
The intersections corresponding to the quark
doublets ($i=-1,0,1$), 
up type singlets ($j=-1,0,1$) and down type singlets ($j^\ast=-1,0,1$)
are denoted by 
an empty circle, full circle and a cross, respectively.
All distances are measured in units
of $2 \pi R$ with $R$ being the corresponding radius (except
$\tilde{\epsilon}^{(3)}$ which is measured 
in units of $6 \pi R$).\label{cap:An-SM-like}}
\end{figure}

Let us now turn to the model. To be specific, we shall concentrate on the
type of set-up shown in figure \ref{cap:An-SM-like}, first introduced
in Ref.~\cite{Cremades:2002va}.
The gauge groups live on stacks of D6 branes, each of
which wraps a 3 one-cycles in $T_{2}\times T_{2}\times T_{2}$.  
Although this set-up is not fully realistic, it captures all 
the features of the relevant contributions to FCNC and Yukawa couplings. 
More realistic 
configurations typically involve D5 and D4 branes (so that the transverse 
volumes can allow a low string scale without diluting the gauge
couplings too much)  
and typically orientifolds. 
The effect of the latter is to introduce mirror branes. 
However, the interactions for strings that end on mirror branes are 
no different to those for strings on the original branes. 
At most, orientifolding
changes gauge groups but makes no difference to the calculation of the 
4 point couplings. What does change with orientifolding is the sums over 
contributions from multiply wrapped worldsheets. However, these contributions
are exponentially suppressed 
and the  calculation of leading terms is the same as 
that for  flat non-compact 
space anyway. 
The situation with 
D4 and D5 branes can be trivially derived from the D6 brane case 
by simply switching off  irrelevant contributions  
to the classical instanton action. (The quantum part generalizes easily but as 
we have said plays a minor role in this discussion.)

In the model of interest, we have four stacks of branes called
baryonic (a), left (b), right (c) and leptonic (d),  
which give rise  
to the gauge groups $U(3)\sim SU(3)\times U(1)_a$, 
$SU(2)$~\footnote{This particular model
  uses the orientifold projection to obtain the gauge group
  $USp(2)\sim SU(2)$ instead of the usual $U(2)$, see
  Ref.\cite{Cremades:2002va}.}, $U(1)_c$ and $U(1)_d$, respectively.  
The matter fields live at the
intersections of the branes and transform as the bi-fundamental
representation of the corresponding groups, so that, 
for example, the open strings
stretched between the U(3) brane and the SU(2) brane have (3,2) quantum
numbers and hence are left handed quarks, the Higgs fields live at
the intersection of the SU(2) and U(1) branes and so on. 
Yukawa couplings correspond to the emission of an open string mode
at, say, the Higgs intersection which then travels to the opposing corners of
a \char`\"{}Yukawa triangle\char`\"{}. 
This is a non-perturbative
process calculable with the help of
conformal field theory techniques, as has recently
been done in Refs.\cite{Cvetic:2003ch,Abel:2003vv,Abel:2003yx}. 
As one might expect, the amplitude 
 is dominated by an exponential of the classical action. 
For 3 point (Yukawa) interactions, the action turns out to 
be equal to the sum of the areas of the triangles projected in each sub-torus.
Thus,
 \[
Y\sim e^{-S_{cl}}\sim e^{-\sum_i \frac{{\rm Area}_i}{2\pi\alpha'}}\]
where $i=1..3$ labels the 2-tori and $\alpha'$ is the string tension. 

For the four (and higher) point couplings, 
there is no such prescription for extracting the contribution 
to the classical action, except for some simple cases \cite{Abel:2003yx}. 
Specifically, for an $N$-point function,
only if the $N$-sided polygons in each sub-torus 
are either zero or the same (up to an overall 
scaling), is the classical action the sum of the projected areas. 
Otherwise factorizability is lost.
We shall see this explicitly in the case of  the four point functions. 
We will also observe that  
the 4 point diagram with a ``Higgs intersection'' reduces 
to the $s$- or $t$-channel Higgs exchange.

\subsection{Generalities of the 4 point function calculation}

\begin{figure}
\begin{center}\includegraphics[%
  scale=0.35]{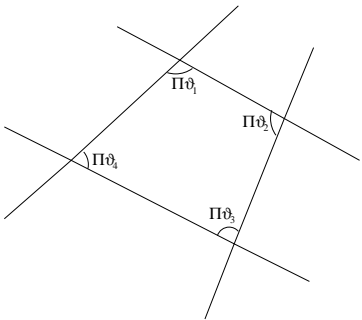}\end{center}
\caption{Generic 4 point string scattering diagram.\label{cap:Generic-4-point}}
\end{figure}

In order to be completely general, we will consider the 
four fermion scattering
amplitude in cases where independent branes intersect at arbitrary
angles 
(bearing in mind that there can be three independent angles 
for $\bar q q \bar l l$ type diagrams). We need to determine the 
instanton contribution shown in figure
\ref{cap:Generic-4-point}. 
If the four intersecting branes 
form the boundary of a convex 4-sided polygon, with interior angles
$\vartheta_{i}$ (in the units of $\pi$), then
\begin{equation}
\sum_{i=1}^{4}\vartheta_{i}= 2. 
\end{equation}
The leading contribution to the amplitude comes from the action 
with least area, which is the same in toroidal or planar cases.  Due to
our choice of the compactification manifold $T^2 \times T^2 \times T^2$,  
the amplitude can be factorized into a product of three
contributions, one from each of the three sub-torus factors. 

Denoting the spacetime coordinates for a 
particular sub-torus by $X=X^{1}+iX^{2}$
and $ \bar{X}=X^{1}-iX^{2}$. The bosonic
field $X$ can be represented as a sum of   a classical piece, $X_{cl}$, and a quantum
fluctuation, $X_{qu}$. The amplitude then
factorizes into classical and quantum components,
\begin{equation}
Z=\sum_{\langle X_{cl} \rangle}e^{-S_{cl}}Z_{qu},
\end{equation}
where
\begin{equation}
S_{cl}=\frac{1}{4\pi \alpha'} \int d^{2}z ~(\partial X_{cl}
  \bar{\partial}\bar{X}_{cl}+\bar{\partial} X_{cl} \partial \bar{X}_{cl})
.
\end{equation}
Here $\alpha'$ denotes the string tension.
For our purposes, only the classical part of the amplitude,
which takes account of the string instantons, is important.
$X_{cl}$ must satisfy the string equations of motion and possess the correct
asymptotic behaviour near the polygon vertices 
(i.e. it has to ``fit'' in the vertex). The task of finding the solutions 
that meet these criteria forms a large part of the 
analyses in Refs.\cite{Cvetic:2003ch,Abel:2003vv,Abel:2003yx}. Having
found the correct 
classical solutions  $X_{cl}$, one then calculates the corresponding
action. The Euclidian action leads to the well known area--suppression
in the amplitude, as expected from instanton considerations.

\begin{figure}
\caption{\label{qqll}$t$-channel Higgs exchange as a {}``double instanton''.}
\begin{center}\includegraphics[%
  scale=0.75]{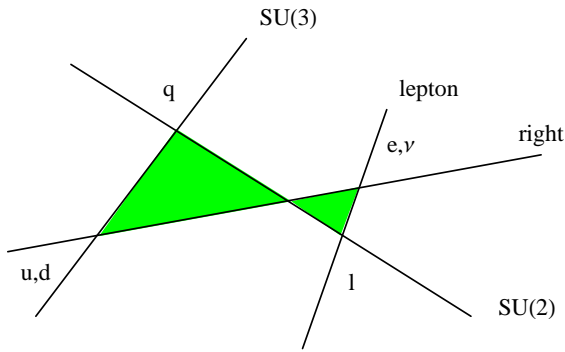}\end{center}
\end{figure}

A more detailed discussion of the calculation of the 4 point function 
are provided in the appendices.
An important check is that the double instanton calculation
agrees with the field theory result.
In particular, the interaction $(\bar u u)(\bar e e)$
produced by combining the two Yukawa interactions as in Fig.\ref{qqll}
goes as
\begin{eqnarray}  
&& {Y_u Y_e \over t-M_H^2} ~, \nonumber
\end{eqnarray}
which is nothing but the field theory $t$-channel Higgs exchange, 
or the $s$-channel equivalent.
However, as mentioned earlier, there are important stringy 
contributions to the same processes that have no field theory equivalent, 
and which are important sources of flavour changing. 
It is to these that 
we now turn.

\subsection{Four point interactions and a model for Yukawa matrices}

\begin{figure}
\caption{\label{qqll2} ``Irreducible'' instanton contribution
to 4 fermion operators.}
\begin{center}\includegraphics[%
  scale=0.75]{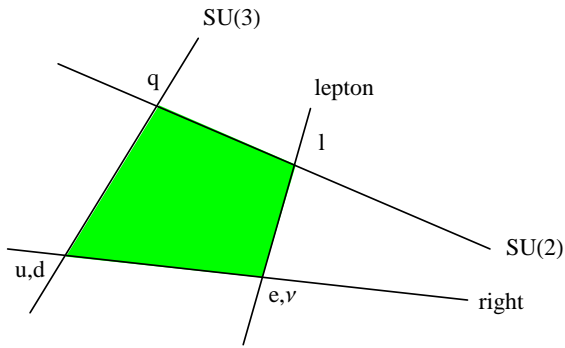}\end{center}
\end{figure}

In string theory, some contact interactions can be generated without
exchanging the Higgs.
They are induced by a single instanton with no field theory
pole (i.e. no brane intersection). It is important to note that such amplitudes 
can be significantly larger than the 
field theory contributions if the string scale is $low$. 
This is because the Higgs exchange involves
the product of two Yukawa couplings. 
It is dominant 
if the leading diagram has one brane on either side of the intersection 
at which the Higgs is located. But if, for example, 
both the $SU(3)$ and lepton branes are lying on the same side of the Higgs 
intersection, as in Fig.(\ref{qqll2}), the contribution to $\bar q_Lq_R\rightarrow \bar e_Le_R$ 
goes roughly as $Y_{e}/Y_{u}$ and can be significantly enhanced for low
string scales\footnote{This is at the moment a heuristic 
argument; the 
Yukawa couplings are really matrices and, as we shall see, the generic 4 point 
instanton coupling does not simply reproduce an inverse Yukawa coupling.}.
In the (unrealistic) limit that the lepton brane is
lying on top of the SU(3) brane in all $T_{2}$ tori, there is
no Yukawa suppression at all in this process. Note, however, that there
should be an overall stringy suppression as there is no field theory
limit and, therefore, no pole. Thus, one expects a contribution of
the type
\[
\frac{Y_{e}/Y_{u}}{M_{S}^{2}} ~. \]

Let us now consider the tree level Yukawa structure in the 
set-up of Fig.\ref{cap:An-SM-like}.             
In this configuration, 
the generation number for the left--handed species varies in one 
of the $T^2$ sub-tori, while that for the 
right--handed species varies in some other sub-torus. 
We will
refer to these tori as the {}``left'' and {}``right'' tori, respectively.
Since the string action is a sum of the $T^2$--projected areas,
the Yukawa coupling contains a factor that depends on the ``left''
generation number only and another factor depending on the 
``right'' generation number, 
\begin{equation}
(Y_{q})_{ij}=a_{i} b_{j} \;.
\end{equation}
This leads to two massless eigenstates, 
which is the ``trivial Yukawa'' problem alluded to earlier.
Clearly, the contact interactions generated by a Higgs exchange 
are also factorizable 
since they are proportional to  a  product of  the Yukawa couplings.
However, generically the stringy 4 point couplings 
induce terms that do not factorize. This will be important for 
us since a nonfactorizable correction to a factorizable structure
generally makes all of the eigenstates massive. This is in contrast to
a factorizable correction, 
$ (Y_{q})_{ij}=a_{i} b_{j} + c_i d_j \;,$
in which case one of the eigenstates remains massless
(this is easily seen by noting that any vector orthogonal to 
$b_j$ and $d_j$ will be annihilated by $Y_{ij}$). In this
case one would need an additional (third) correction of a different
form. In fact, this situation occurs when the 
corrections to the Yukawa couplings are generated by the Kaluza--Klein
mode exchange (which have a nontrivial flavour structure due
to a generation--dependent coupling to KK modes of the gauge bosons
\cite{Carone:1999nz}).
Here we 
concentrate on the  stringy instanton contributions 
in which case realistic flavour structures can be
obtained. 

\begin{figure}
\caption{\label{LRamplitudenocross.eps}
Chirality and flavour changing four fermion string amplitude.
The corresponding polygons are generally {\it non--planar},
even for $i=l$. See also Fig.(\ref{cap:An-SM-like}).}
\begin{center}\includegraphics[%
  scale=0.7]{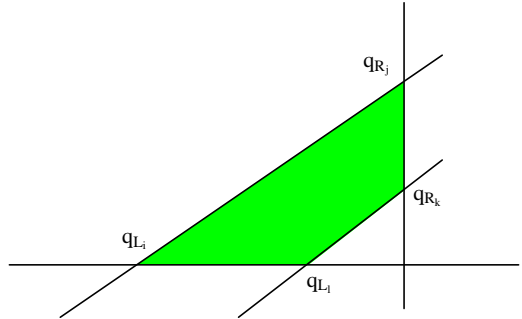}\end{center}
\end{figure}

Let us proceed by considering first a simple case when
the relevant intersection quadrangle has a non--zero
projection in one of the $T^2$ sub-tori only.
By chirality considerations,
in order to induce a correction to the Yukawa coupling,
 we need an operator
of the type $(\overline{q}_{L_{i}}q_{R_{j}})(\overline{q}_{R_{j'}}q_{L_{i'}})$
in Fig.(\ref{LRamplitudenocross.eps}),
which we discuss in more detail in Sec.4.3.
The corresponding contribution to the 
$(\overline{q}_{L_{i}}q_{R_{j}})(\overline{q}_{R_{j'}}q_{L_{i'}})$
amplitude involves \cite{Abel:2003yx} (see also the appendices)
\begin{equation}
S_{cl}=\frac{1}{2\pi\alpha'}
\left(\frac{\sin\pi\vartheta_{1}\sin\pi\vartheta_{4}}
{\sin(\pi\vartheta_{1}+\pi\vartheta_{4})}
\frac{v_{14}^{2}}{2}+\frac{\sin\pi\vartheta_{2}
\sin\pi\vartheta_{3}}{\sin(\pi\vartheta_{2}
+\pi\vartheta_{3})}\frac{v_{23}^{2}}{2}\right)\;.
\end{equation}
Here $\vartheta_i$ are the angles and $v_{14}, v_{23}$ are the 
sides of the quadrangle. 
Noting that 
$\sin(\pi\vartheta_2+\pi\vartheta_3)=-\sin(\pi\vartheta_1+\pi\vartheta_4)$,
one may verify that this is simply
the area/$2\pi\alpha'$ of the four--sided polygon. 
If the generation number $i$ of the left--handed quarks
varies in this sub-torus, while that of the right--handed 
quarks $j$ varies in an orthogonal torus, the result is
independent of $j$ and the amplitude is proportional to
$Y_{ij}/Y_{i'j}=a_{i}/a_{i'}$.
This means that the 4 point function factorizes into a
``left--handed'' piece times a ``right--handed'' piece. 
Factorizability is also found in the ``degenerate'' case 
when the polygons 
in each $T_2$ torus 
are equivalent up to an overall scaling. 
This is because the 
action is again simply the sum of the projected areas.

In a more general situation, the classical action is no longer
given by the sum of the projected areas. This is due to the
presence of two conflicting contributions in the action
such that its minimization does not produce a factorizable result.
Consider the simplest non-trivial case, which
is when the angles are the same for each sub-torus but the lengths $v_{kl}$
differ. As we are interested in the operator 
$(\overline{q}_{L_{i}}q_{R_{j}})(\overline{q}_{R_{j'}}q_{L_{i'}})$,
we have only two independent angles and  $\vartheta_{1}=1-\vartheta_{2}$
and $\vartheta_{4}=1-\vartheta_{3}$.
As we show in the appendices, the contribution to the 
coupling is dominated by a saddle point where the action 
in minimized. This gives \cite{Abel:2003yx}
\begin{equation}
S_{cl}=\frac{1}{4\pi\alpha'}\frac{\sin\pi\vartheta_{2}\sin\pi\vartheta_{3}}{\sin(\pi\vartheta_{2}+\pi\vartheta_{3})}\sqrt{\sum_{m}\left(v_{23}^{m}-v_{14}^{m}\right)^{2}\,\sum_{n}(v_{23}^{n}+v_{14}^{n})^{2}}~,\label{non-crossing}
\end{equation}
where $m,n$ label the three sub-tori.
Only for the trivial or degenerate cases described above does $e^{-S_{cl}}$
factorize into a ``left--handed'' and a ``right--handed'' pieces.
This non--factorizability propagates into the Yukawa matrices
through loop corrections.

Consider the model of  Fig.\ref{cap:An-SM-like}.
For simplicity, we assume 
a common ratio of vertical to horizontal radii 
in the second and third tori: 
$R^{(2)}_2/R^{(2)}_1=R^{(3)}_2/R^{(3)}_1 \equiv \chi$, where
$R^{(I)}_{1,2}$ are the horizontal and vertical radii, respectively,
of the $I-$th torus.
The fact that the left and right branes are at
right angles in the second and third tori and that the 
operator we are interested in is
$(\bar{q}_{L_i} q_{R_j}) (\bar{q}_{R_{j^\prime}} q_{L_{i^\prime}})$
leaves us with just one independent angle for all of the 
amplitudes, namely $\theta^{(2)}_{ab}=\frac{1}{2}-\theta^{(2)}_{ac}
=\theta^{(3)}_{ac}=\frac{1}{2}-\theta^{(3)}_{ab}\equiv \sigma=
\frac{1}{\pi}\tan^{-1}(3\chi)$.
The location of the branes is 
parameterized by $\epsilon_2$, $\epsilon_3$ and
$\tilde{\epsilon}_3$ as shown in Fig.\ref{cap:An-SM-like}.
There
are two different types of diagrams that can contribute, depending
on whether there is a crossing of the branes, so that the area swept out
by propagating strings
is formed by two triangles joined by the Higgs vertex, or there is no
crossing, so that the relevant area is a convex four-sided polygon. 
If the areas in the two sub-tori 
are of the same kind, 
the corresponding action can be minimized analytically. 
For diagrams with a crossing, the result is proportional
to the Yukawa couplings and no new flavour structure emerges.
For diagrams without a crossing,
the  action is given by
Eq.(\ref{non-crossing}) and
the corresponding correction to the Yukawa vertex is non--factorizable. 
The
mixed case, crossing in one torus and non-crossing in the other,
is more involved, depending on which 
sub--torus gives a dominant contribution to the action.
If it is the non-crossing one, then the
amplitude still factorizes, although it is no longer proportional to
the Yukawa couplings (or the projected areas). 
It introduces  new non-trivial, although factorizable, 
flavour structure. On the other hand, if the crossing diagram dominates, 
again a non-factorizable correction is found.

This new flavour structure
appearing in the chirality changing four-fermion amplitude will
propagate through radiative corrections to the Yukawa couplings. Using
this amplitude as an effective vertex, we can estimate the one loop
threshold correction to the Yukawa couplings, leading to a generic
form
\begin{equation}
Y_{ij} \approx a_i b_j + \frac{\alpha}{\pi} A^{LR}_{ijkl} a_k b_l ,
\end{equation}
where $A^{LR}_{ijkl}$ is the coefficient 
 of the operator $ \bar{q}_{L_i} q_{R_j} \bar{q}_{R_k} q_{L_l}$
and $\alpha/\pi$ represents a loop suppression.  

This form ensures a natural hierarchy between the third and the first
two families, with their masses generated at tree and one loop level,
respectively.
In order
to get some intuition about the hierarchical structure of the first
two families, it is useful to consider the limit in which the new 
contribution almost factorizes (recall that there are factorizable
corrections that are not proportional to the Yukawa couplings) 
with a small non-factorizable correction
\begin{equation}
Y^{u,d}_{ij}=a_i b^{u,d}_j+ \frac{\alpha}{\pi} 
\big(c_i d^{u,d}_j + \epsilon \tilde{C}^{u,d~LR}_{ij}\big),
\label{perturbative:yuk}
\end{equation}
where $\epsilon$ measures departure from factorization in  the
chirality changing four-fermion amplitude.
In the limit $\epsilon \rightarrow 0$, there is
a massless state since there exists a 3D vector orthogonal
to $b_j$ and $d_j$ and therefore annihilated by $Y_{ij}$.
The matrices $Y_{ij}^{u,d}$ can be  diagonalized  perturbatively leading to the following
values of the diagonal Yukawa couplings
\begin{equation}
Y^{u,d}_1= \frac{\alpha}{\pi} \epsilon \mu^{u,d}_{11},
\quad
Y^{u,d}_2=\frac{\alpha}{\pi} \mu^{u,d}_1,
\quad
Y^{u,d}_3= |a||b^{u,d}|,
\end{equation}
and the mixings
\begin{align}
V^{CKM}_{12}=&
\epsilon \Big[ \frac{\mu^d_{12}}{\mu^d_1}-\frac{\mu^u_{12}}{\mu^u_1}\Big],
\nonumber \\
V^{CKM}_{13}=&
\frac{\alpha}{\pi} \epsilon \frac{1}{|a|}
\Big[
\frac{\mu^{u}_{12}\mu^{u}_{2}/\mu^{u}_{1}-\mu^{u}_{13}}
{|b^{u}|}
-
\frac{\mu^{u}_{12}\mu^{d}_{2}/\mu^{u}_{1}-\mu^{d}_{13}}
{|b^{d}|}
\Big],
\nonumber \\
V^{CKM}_{23}=&
\frac{\alpha}{\pi} \frac{1}{|a|} \Big[
\frac{\mu^d_2}{|b^d|}-\frac{\mu^u_2}{|b^u|}
\Big].
\nonumber
\end{align}
Here
$\mu_i^{u,d}$ and $\mu_{ij}^{u,d}$ are order one functions
of $a_i,b_i,c_i,d_i$ and $\tilde{C}^{LR}_{ij}$.
The hierarchical pattern of quark masses and mixing angles found
in nature~\cite{PDG}
\begin{align}
&m_u\sim 3\times 10^{-3} \mbox{ GeV},\quad  m_c\sim 1.2\mbox{ GeV},
\quad  m_t\sim 174 \mbox{ GeV}, \nonumber \\
&m_d\sim 7\times 10^{-3} \mbox{ GeV},\quad m_s\sim 0.12\mbox{ GeV},
\quad  m_b\sim 4.2 \mbox{ GeV},\label{massesandmixings:data}\\
& V_{12}\sim 0.22, \quad V_{13}\sim 0.0035, \quad V_{23}\sim 0.04,
\nonumber
\end{align}
can be explained by a hierarchy in the expansion coefficients,
$\alpha$ and $\epsilon$. In fact, reasonable values for all experimental
data in Eq.(\ref{massesandmixings:data})
can be obtained with 
\begin{equation}
\frac{\alpha}{\pi}\sim 10^{-2}, \quad \epsilon \sim 0.1,
\end{equation}
except for the up quark for which some amount of cancellation seems
necessary.

In the model of Fig.\ref{cap:An-SM-like} 
things are a bit more involved, 
but a reasonable estimate can still be obtained.
Keeping the ratio of the vertical to horizontal radii the same in the
second and third tori (\textit{i.e.} just one $\chi$), the same $N=1$
supersymmetry is preserved at all the intersections. In that case, the
threshold correction from the four point amplitude 
vanishes due to non-renormalization theorems. As we shall see in
the next sections, there are still sources of non-trivial Yukawa
matrices even in that case. 
Consider now
 $\chi_2\neq \chi_3$.
The relevant
parameters determining the flavour structure are then the horizontal radii
of the second and third tori, $R^{(2,3)}_1$, the vertical to
horizontal radii ratios, $\chi_{2,3}$ and the locations of the branes
parameterized by $\epsilon_2$, $\epsilon_3$ and $\tilde{\epsilon}_3$
(see Fig.~\ref{cap:An-SM-like}). Complex phases appear
due to the Wilson lines or the antisymmetric backgroud
field~\cite{Cremades:2002va}. 
These are however irrelevant for
the tree level Yukawa couplings since the factorization makes it
possible to rephase them away. 
Non-trivial backgrounds   also
generate complex phases 
in the four-fermion amplitudes which
then propagate through the threshold effects to the Yukawa couplings. 
We  estimate their
effects by adding random order one phases to the different entries of
the amplitudes. 
The final parameter necessary to compute the quark 
 masses is the ratio of the two Higgs VEVs, $\tan \beta$.
For the following values of the parameters (dimensionful
parameters are in string units)
\begin{align}
&R^{(2)}_1=1.1, \quad
R^{(3)}_1=1.15, \quad
\chi_2=1.24,\quad
\chi_3=0.94,\quad
\nonumber \\
&\epsilon_2=0.121,\quad
\epsilon_3=0.211,\quad
\tilde{\epsilon}_3=0.068,\quad
\tan\beta=20, \label{our:parameters}
\end{align}
the spectrum of quark masses and mixing angles is reproduced 
with reasonable accuracy (recall a factor of a few uncertainty
in the light quark masses):
\begin{align}
&m_u\sim 4\times 10^{-3} \mbox{ GeV},\quad  m_c\sim 1.8\mbox{ GeV},
\quad  m_t\sim 176 \mbox{ GeV}, \nonumber \\
&m_d\sim 4\times 10^{-3} \mbox{ GeV},\quad m_s\sim 0.04\mbox{ GeV},
\quad  m_b\sim 8 \mbox{ GeV},\label{massesandmixings:ourresult}\\
& V_{12}\sim 0.22, \quad V_{13}\sim 0.003, \quad V_{23}\sim 0.02,
\quad J={\cal O}(10^{-5}),
\nonumber
\end{align}
where we have included a global normalization factor 0.95 in the Yukawa
couplings. 
Although the matching is not perfect,
the  point here 
is that a semirealistic pattern of fermion masses and
mixing angles arises once the non-trivial flavour
structure at one loop  is taken into account. The  rotation
matrices defined by 
\begin{equation}
L_{u,d}^\dagger Y_{u,d} R_{u,d}=Y_{u,d}^{\mathrm{diag}}
\label{L}
\end{equation}
take the following values:
\begin{equation}
|L_u|=\begin{pmatrix}
0.12 & 0.84 & 0.53 \\
0.003 & 0.54 & 0.84 \\
0.99 & 0.10 & 0.07
\end{pmatrix},
\quad
\mathrm{Arg}(L_u)=\begin{pmatrix}
-3.14 & 3.13 & -3.12 \\
-2.50 & -0.02 & -3.13 \\
0 & \pi & \pi
\end{pmatrix},\label{Lumatrices}
\end{equation}
\begin{equation}
|R_u|=\begin{pmatrix}
0.10 & 0.40 & 0.91 \\
0.43 & 0.82 & 0.37 \\
0.90 & 0.40 & 0.17
\end{pmatrix},
\quad
\mathrm{Arg}(R_u)=\begin{pmatrix}
-2.74 & -0.24 & -3.13 \\
1.60 & 2.79 & -3.12 \\
-0.97 & -2.82 & -3.14
\end{pmatrix},\label{Rumatrices}
\end{equation}
\begin{equation}
|L_d|=\begin{pmatrix}
0.13 & 0.83 & 0.54 \\
0.12 & 0.53 & 0.84 \\
0.98 & 0.17 & 0.07
\end{pmatrix},
\quad
\mathrm{Arg}(L_d)=\begin{pmatrix}
-1.47 & 1.94 & -3.13 \\
2.38 & -1.23 & -3.14 \\
0 & 0 & \pi
\end{pmatrix},\label{Ldmatrices}
\end{equation}
and
\begin{equation}
|R_d|=\begin{pmatrix}
0.43 & 0.64 & 0.63 \\
0.31 & 0.56 & 0.77 \\
0.84 & 0.52 & 0.10
\end{pmatrix},
\quad
\mathrm{Arg}(R_d)=\begin{pmatrix}
2.50 & 1.98 & -3.13 \\
-0.34 & -1.28 & -3.14 \\
-1.62 & 0.84 & -3.13
\end{pmatrix}.\label{Rdmatrices}
\end{equation}
These are the matrices we will use in the following sections to
derive the FCNC bounds. Their effect (at least in our example)
is non--trivial as they have significant off--diagonal entries
and thus cannot be approximated by ``a small angle'' rotation
matrix.

Finally, we note that if the low energy theory is described neither
by the SM nor by the MSSM type model, non--trivial flavour structures can
appear at tree level. For example, a model with 6 Higgs doublets has
been studied in Ref.\cite{Chamoun:2003pf}  and has been shown to generate a realistic spectrum.

\subsection{One loop Yukawa thresholds in string theory}

Here we have not attempted to calculate the 1--loop thresholds directly in 
string theory, which we defer until a subsequent publication. 
But there are a number of comments to make in this regard. In particular,
it is interesting to see how the non-renormalization theorem appears in 
the string calculation.

\begin{figure}
\begin{center}\includegraphics[%
  scale=0.4]{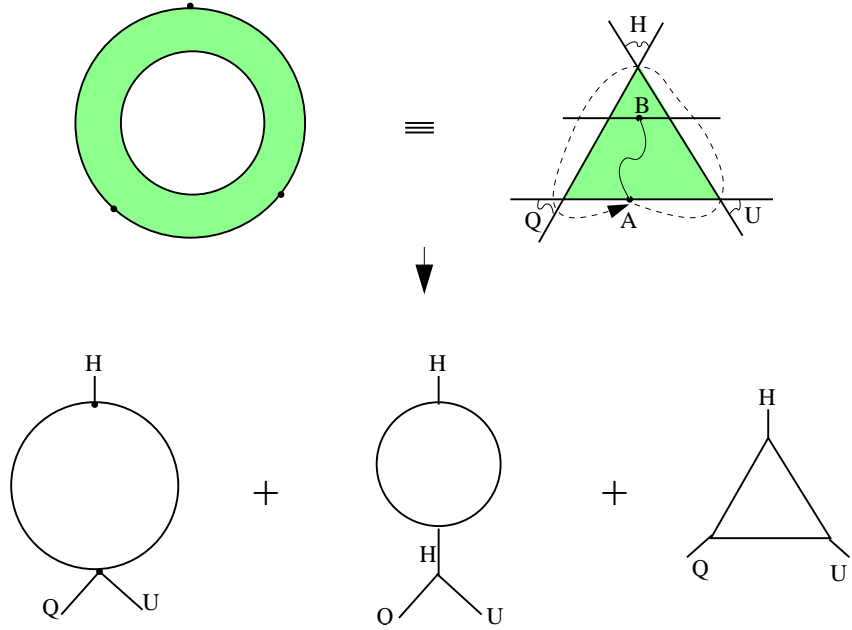}\end{center}
\caption{\label{annulus}Threshold correction to Yukawa
couplings in string theory. The string diagram is an annulus with three 
twist vertices on the external boundary. In the target space, 
the diagram corresponds to taking an open string stretched between two branes as shown, 
and moving one end around the Yukawa triangle. Various limits correspond to either the 
field theory threshold diagram with a four point operator inserted or 
the usual field theory renormalization diagrams. The wave function
renormalization should be all that 
remains in supersymmetric configurations.}

\end{figure}

The relevant diagrams are shown in figure \ref{annulus}. The annulus diagram 
has three vertex operators inserted on one boundary corresponding to the 
$H,Q,U$ fields. These operators include bosonic twist fields whose job is to 
take the end of one string and move it from one brane to the next. The diagram 
therefore corresponds to the physical situation shown, where we have a 
string stretched between two branes. When we keep one end (B) fixed on some 
brane and move the opposing end (A) around the Yukawa triangle, we generate 
the three states appearing in the Yukawa coupling. The end (B) remains on the 
same brane and forms the inside edge of the annulus. It can be 
attached to any brane, although contributions from branes at large 
distance from the Yukawa triangle are suppressed by instanton 
contributions. From a field theory point of view these would 
correspond to heavy stretched string states propagating in the loop.
The diagram shows the case where end B is on the $SU(3)$ brane
where the fields propagating in the 
loop are quarks and gluons. If B is on the $SU(2)$ brane, then 
the fields in the loop 
are the $W$ boson, Higgs and left-handed quark. Note that the $W$ boson is 
an ``untwisted'' state so that one would expect a sum over $W$ 
Kaluza-Klein modes from these diagrams. 

There are different field theory limits that one can take when evaluating 
the instanton action, which are also shown in the figure. 
Labelling the positions of the vertices on the 
boundary as $x_H$, $x_Q$, $x_U$, 
the one loop field theory diagrams shown are extracted
in the limit $x_Q \rightarrow x_U$ 
(Note that only one of these is fixed by residual symmetry). 
We then have two possibilities. 
The first case corresponds to an instanton four point vertex and a Yukawa 
coupling. (The area swept out by the instanton world-sheet will 
be roughly the large triangle which is the sum of the small triangle and the
four point area, so heuristically this makes sense. In 
$e^{- S_{cl}}$ we would expect to get a product of the two 
couplings.) 

The second case occurs when the ``A'' endpoint travels 
through the Higgs intersection sweeping out three Yukawa triangles 
(or something approximating that). This corresponds to the 
standard Yukawa coupling renormalization.

The threshold contributions 
should decouple in supersymmetric theories, where 
only the field renormalization contributions are present
by the non-renormalization theorem. The cancellation comes
from a prefactor in the Yukawa couplings. 
The prefactor is found by factorizing the 
amplitude on the one loop 
partition function when all the vertices come together. 
In this limit one is left with the partition function on the 
annulus with ends on the two relevant branes. If these branes 
are tilted then supersymmetry can be completely broken, 
otherwise the prefactor vanishes by the ``abstruse identity'' if the 
branes are parallel. In this way we can see that in order to 
get a non-zero threshold in cases where the visible states preserve $N=1$ 
supersymmetry, requires the interior of the annulus (end B) to be 
on a hidden brane. The threshold contributions will therefore come 
only from non-susy diagrams that involve states stretched between 
visible sector branes and the susy breaking hidden branes. 

In the $N=1$ case, one has to be a little more 
careful because not all contributions
vanish, and those infra-red divergences that correspond to 
field renormalization should remain. This behaviour 
has to do with the Higgs pole term which is present for the 
field renormalization terms but not for the threshold terms. 
One expects (although this has to be checked) 
that the only non-vanishing contributions to 
these diagrams in supersymmetric theories are proportional to a 
factor $t-m^2$, so that, on mass-shell, 
the only non-vanishing pieces are those with a Higgs pole.

\section{Remarks on flavour in supersymmetric models}

We have shown that rich flavour structures arise in non--supersymmetric
intersecting brane models at loop level.
However, as we saw in the previous section, 
these  arguments do not apply to globally supersymmetric models.
This is due to non--renormalization of the N=1 superpotential
which forces any  
 threshold corrections to the Yukawa couplings 
to be suppressed by $m_{\rm SUSY}^2/M_S^2$,
where $m_{\rm SUSY}$ is the soft breaking mass. 
The resulting quark masses are far too small.
Also, we note that the brane intersection angles are
fixed by supersymmetry, so there is less freedom
in choosing the desired parameters.

Nevertheless, there is an additional source of flavour structures
in supersymmetric models. In string theory, the soft breaking A--terms,
although related to the Yukawas, generally have a different
flavour pattern.
This property is desirable from the phenomenological perspective
and allows one to derive constraints on models of flavour
and/or SUSY breaking  \cite{Abel:2001cv}.
Specifically, spontaneous breaking of supergravity requires 
\cite{Soni:1983rm}
\begin{equation}
A_{\alpha\beta\gamma}=F^m \left[
\hat K_m + \partial_m \ln Y_ {\alpha\beta\gamma} -
\partial_m \ln (\tilde K_\alpha \tilde K_\beta \tilde K_\gamma)
\right].
\end{equation}
Here the Latin indices refer to the ``hidden sector'' fields,
while the Greek indices refer to the observable fields;
the K\"ahler potential is expanded in observable fields as 
$K= \hat K + \tilde K_{\alpha} \vert C^\alpha \vert^2 +...$
and $\hat K_m \equiv \partial_m \hat K$. The sum in $m$
runs over SUSY breaking fields. 
The index convention is $Y_{H_1 Q_i D_j} \equiv Y^d_{ij}$
and so on. The soft trilinear parameters enter in the 
soft breaking Lagrangian as
\begin{equation}
\Delta {\cal L}_{\rm soft}= -{1\over 6} A_{\alpha\beta\gamma} 
Y_{\alpha\beta\gamma} C^\alpha C^\beta C^\gamma \;.
\end{equation}
An analysis of SUSY soft breaking terms in intersecting brane models
has recently been performed in \cite{Kors:2003wf}.

What is important for us is that the A--terms are always flavour--dependent
unless the moduli entering the Yukawa couplings do not break supersymmetry
\cite{Abel:2001cv},
\begin{equation}
\Delta A_{ij} =   F^m \partial_m \ln (a_i b_j) \;,  
\end{equation}
where $Y_{ij}=a_i b_j$.
The Yukawa couplings depend, first of all, on the compactification
radii, so the relevant moduli are the $T$--moduli. These generally
break SUSY (unless a special Goldstino angle is realized).
As a result, the soft breaking Lagrangian will contain a new
flavour structure, not proportional to the original Yukawa matrices,
\begin{equation}
\Delta {\cal L}_{\rm soft}\sim  {\rm const} 
\sum_{ij} a_i b_j ( c_i + d_j) ~\tilde q_{L_i}^* \tilde q_{R_j} H +... 
\end{equation}

\begin{figure}
\caption{\label{Yuksusy.eps}
SUSY correction to the Yukawa coupling.}
\begin{center}\includegraphics[%
  scale=1]{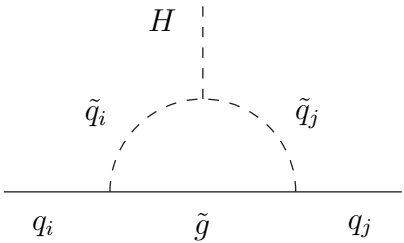}\end{center}
\end{figure}

The SUSY vertex corrections to the Yukawa interactions
modify the tree level quark masses and mixings \cite{Hempfling:1993kv}. 
In particular,
the gluino mediated diagram of Fig.\ref{Yuksusy.eps}
generates  a  correction  
\begin{eqnarray}
&&Y_{ij}^{\rm } = a_i^{\rm } b^{\rm }_j + 
 \epsilon~ a^{\rm }_i b^{\rm }_j (c^{\rm }_i + d^{\rm }_j) +...
\end{eqnarray}
and analogously  for the chargino and neutralino contributions.
Note that this contribution does not decouple as the string scale
or the soft breaking scale is raised.
Here we have assumed universal masses for the sfermions in the loop.
Corrections of this form induce masses for the second generation
since there is only one 3D vector 
orthogonal to both $b^{\rm }_j$ and  $b^{\rm }_j d^{\rm }_j $
corresponding to a massless eigenstate.
More sophisticated contributions (which include the RG running of
the soft masses in the loop) can make all of the eigenstates massive.
Whether this mechanism indeed leads to a realistic spectrum requires a
separate study.

\section{Constraints from flavour and CP physics}

In this section, we analyze constraints on the string scale 
due to flavour and CP physics. The main point is that
the mechanism of family replication in intersecting brane models
leads to the existence of a large class of four fermion 
flavour and CP violating operators. These are suppressed by
the string scale squared such that the experimental 
constraints can be interpreted as constraints
on the string scale. In what follows, we first consider
flavour violating processes and then turn to the 
flavour conserving CP--violating observables, the EDMs.

\subsection{FCNC bounds}

\begin{figure}
\caption{\label{LLamplitude.eps}
Chirality--preserving flavour changing string amplitude. See also Fig.(\ref{cap:An-SM-like}).}
\begin{center}\includegraphics[%
  scale=0.6]{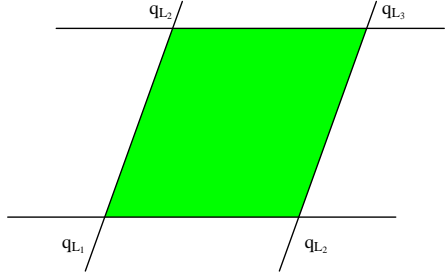}\end{center}
\end{figure}

Instanton--induced 4 point amplitudes have allowed us to
obtain a realistic pattern of fermion masses and mixing angles 
even in
the simplest models with intersecting branes. They also contribute to
{\it tree level} flavour violating transitions that are much suppressed in
the Standard Model and 
extremely well constrained by experiment. 
Some of them (in particular, chirality conserving operators of 
Fig.\ref{LLamplitude.eps})
along with FCNC generated by the exchange of gauge boson KK modes
were considered in Ref.\cite{Abel:2003fk}.
It was observed that the two sources of FCNC are
$complementary$ in their dependence on
the compactification radii, such that
an absolute bound  $M_S \gtrsim 100$ TeV can be obtained
independently of the ``size'' of the extra dimensions.
 As we have
emphasized in the introduction, this bound can only be treated as an estimate
given the fact that   a realistic theory of flavour was absent at that time.
 Now that we have one at our disposal we can
make more reliable 
predictions in flavour physics and derive robust bounds on the string scale.
Our mechanism of generating flavour structures fixes the compactification
radii to be of order the string length because,  
in order to get
enough flavour at one loop, the tree level Yukawa matrices must
have relatively large entries.
As a result, the bounds we obtain now
are stronger than those derived previously.

It should be noted that the bounds from flavour physics generally depend
on the flavour model. In particular, we have succeeded in generating realistic flavour structures
from instanton--induced operators. One may also attempt to derive these from
flavour--dependent quark couplings to the gauge KK modes
which are important for large compactification radii. In
 this case the tree level Yukawa matrices are strongly hierarchical.
We find that the corresponding flavour structures are quite restrictive
and a realistic spectrum cannot be obtained (at least at one loop).

In this subsection, we  discuss the most important flavour violating
observables and the constraints they impose.
We  closely follow the analysis of
Ref.~\cite{Langacker}, where a phenomenological study of models
with a flavour non-universal extra $\mathrm{U}(1)$ was performed.
The effect of the $Z'$ exchange is mimicked by string instantons
in our model. 
The
four-fermion amplitudes are denoted by $A^{\chi_1 \chi_2}_{ijkl}$,
where $\chi_{1,2}$ are the chiralities of the amplitude and
$i,j,k,l$ are the generation indices. As the calculation is performed 
in the physical basis, we have
to include the relevant rotation matrices. 
For instance,
the ``left-left'' amplitude with four up type quarks reads
\begin{equation}
A^{LL}_{u_au_bu_cu_d}=\sum_{ijkl} A^{LL}_{u_iu_ju_ku_l} 
(L^\dagger_u)_{ai} (L_u)_{jb} (L^\dagger_u)_{ck} (L_u)_{ld},
\end{equation}
where $a,b,c,d$ are flavour indices
and $L_u$ is defined by (\ref{L}). A similar expression holds for
the other amplitudes.

We start our discussion with observables in the quark sector 
which we calculate using the rotation matrices 
(\ref{Lumatrices}).
Leptonic and semileptonic
observables will be  $estimated$ using rotation matrices with small
angles.

\subsubsection{Quark sector observables}

The most constraining  observables in the quark sector are those
related to meson oscillations. In the SM, they are induced by one loop
Cabibbo-suppressed box diagrams 
forcing them to be very small.
The extreme experimental accuracy in oscillation
measurements makes them
an ideal place to constrain new flavour physics.

In our case,
both the chirality preserving and chirality changing four-fermion
amplitudes mediate meson oscillations at tree level. The mass
splitting for a meson with quark content $P_0=\bar{q}_j q_i$, in the
vacuum insertion approximation, reads
\begin{equation}
\Delta m_P = \frac{2 m_P F_P^2}{M_S^2}
\Big \{\frac{1}{3} \mathrm{Re}\big[ A^{LL}_{ijij} + A^{RR}_{ijij}
  \big]
-\Big[ \frac{1}{2}+\frac{1}{3} \Big(\frac{m_P}{m_{q_i}+m_{q_j}}
\Big)^2
\Big] \mathrm{Re} A^{LR}_{ijij} \Big\},
\end{equation}
where $m_P$ and $F_P$ are the mass and decay constant of the meson,
respectively. Here $A_{ijkl}$ are the dimensionless coefficients
parametrizing the 4f operators (with $1/M_S^2$ factored out).

Indirect CP violation in the Kaon system, 
which has been measured with  extreme accuracy, 
is parametrized by
\begin{equation}
|\epsilon_K|=
\frac{m_K F_K^2}{\sqrt{2} \Delta m_K M_S^2}
\left|
\Big \{\frac{1}{3} \mathrm{Im}\big[ A^{LL}_{dsds} + A^{RR}_{dsds}
  \big]
-\Big[ \frac{1}{2}+\frac{1}{3} \Big(\frac{m_K}{m_{d}+m_{s}}
\Big)^2
\Big] \mathrm{Im} A^{LR}_{dsds} \Big\}
\right|.
\end{equation}

Numerically,
the experimental constraints 
impose the following bounds:
\begin{itemize}
\item Kaon mass splitting
\begin{equation}
\frac{1}{M_S^2} 
\left|
\mathrm{Re}\big[ A^{LL}_{dsds} + A^{RR}_{dsds}
  \big]
-17.1\, \mathrm{Re}\big[ A^{LR}_{dsds} \big]
\right|
\lesssim 3.3 \times 10^{-7} \mbox{ TeV}^{-2},
\end{equation}
\item $B$ mass splitting
\begin{equation}
\frac{1}{M_S^2} 
\left|
\mathrm{Re}\big[ A^{LL}_{dbdb} + A^{RR}_{dbdb}
  \big]
-3\, \mathrm{Re}\big[ A^{LR}_{dbdb} \big]
\right|
\lesssim 2 \times 10^{-6} \mbox{ TeV}^{-2},
\end{equation}
\item $B_s$ mass splitting
\begin{equation}
\frac{1}{M_S^2} 
\left|
\mathrm{Re}\big[ A^{LL}_{sbsb} + A^{RR}_{sbsb}
  \big]
-3\, \mathrm{Re}\big[ A^{LR}_{sbsb} \big]
\right|
\lesssim 6.6 \times 10^{-5} \mbox{ TeV}^{-2},
\end{equation}
\item $D$ mass splitting
\begin{equation}
\frac{1}{M_S^2} 
\left|
\mathrm{Re}\big[ A^{LL}_{ucuc} + A^{RR}_{ucuc}
  \big]
-3.9\, \mathrm{Re}\big[ A^{LR}_{ucuc} \big]
\right|
\lesssim 3.3 \times 10^{-6} \mbox{ TeV}^{-2},
\end{equation}
\item Kaon CP violation
\begin{equation}
\frac{1}{M_S^2} 
\left|
\mathrm{Im}\big[ A^{LL}_{dsds} + A^{RR}_{dsds}
  \big]
-17.1\, \mathrm{Im}\big[ A^{LR}_{dsds} \big]
\right|
\lesssim 2.6 \times 10^{-9} \mbox{ TeV}^{-2},
\end{equation}
\end{itemize}
It is clear  that if the coefficients of the 4 fermion operators
are order one (times $1/M_S^2$), the Kaon system  puts a constraint 
$M_S \gtrsim 10^{3-4}$ TeV. Later in this section we will use 
 specific values of our rotation matrices to obtain more precise 
bounds.

\subsubsection{(Semi) leptonic observables}

We now turn to leptonic and semileptonic observables. Although some of
them are very well determined experimentally, 
 they only allow
us to obtain a rough estimate of $M_S$ since we do not 
properly address the issues of lepton flavour in this paper.
The most constraining observables
 are the coherent $\mu-e$
conversion in atoms, heavy lepton decays and (semi)leptonic meson
decays. Radiative lepton decays, such as $\mu \to e \gamma$, which are
very important modes for some models of new physics (e.g. supersymmetry), 
are one loop transitions and, therefore, less restrictive than
the tree level ones. 

The expression for the new contribution to $\mu-e$ conversion is long
and not very illuminating. Since we are only interested in an estimate,
we  consider the contribution of  the
left handed fields only. In that case, the  muon conversion leads to the
following constraint on the amplitude
\begin{equation}
\frac{1}{M_S^2} \left|
A^{LL}_{e \mu  uu} + 1.1 A^{LL}_{e\mu dd}  
\right|
\leq 1.1 \times 10^{-6} \mbox{ TeV}^{-2},
\end{equation}
which imposes a bound on $M_S$ of roughly $10^{2-3}$ TeV.

Next, consider tau decays into
three electrons or three muons. The contribution of the string
instantons to the decay width is
\begin{equation}
\Gamma(l_j \to l_i l_i \bar{l}_l)=\frac{m_{l_j}^5}{384 \pi^3 M_S^4}
\Big(
2\big| A^{LL}_{l_il_j l_il_i}\big|^2
+2\big| A^{RR}_{l_il_j l_il_i}\big|^2
+\big| A^{LR}_{l_il_j l_il_i}\big|^2
+\big| A^{RL}_{l_il_j l_il_i}\big|^2
\Big).
\end{equation}
Using the experimental bounds on $\tau \to 3 e$ and $\tau \to 3
\mu$, respectively, we obtain  the following constraints on the string
amplitudes 
\begin{equation}
\frac{1}{M_S^4}
\Big\{
\big| A^{LL}_{e \tau ee}\big|^2
+0.78\big| A^{RR}_{e \tau ee}\big|^2
+0.39\big| A^{LR}_{e \tau ee}\big|^2
+0.5\big| A^{RL}_{e \tau ee}\big|^2
\Big\}
\leq 2.3 \times 10^{-4}\mbox{ TeV}^{-4},
\end{equation}
and
\begin{equation}
\frac{1}{M_S^4}
\Big\{
\big| A^{LL}_{\mu \tau \mu\mu}\big|^2
+0.78\big| A^{RR}_{\mu \tau \mu\mu}\big|^2
+0.39\big| A^{LR}_{\mu \tau \mu\mu}\big|^2
+0.5\big| A^{RL}_{\mu \tau \mu\mu}\big|^2
\Big\}
\leq 1.2 \times 10^{-4}\mbox{ TeV}^{-4}.
\end{equation}
These are inferior to the muon conversion constraints.

Flavour violating four fermion operators induce tree level
corrections to leptonic and semileptonic decays of pseudoscalar
mesons. The complete expressions are again long, complicated functions
of the four-fermion coefficients, and in order to estimate the bounds
we  neglect the mixed LR contributions. Of the multitude
of rare meson decays that have been measured, only the $K^0_L$ decays
to two muons or two muons plus a pion are known with sufficient
experimental precision to put strict bounds on new physics effects.
The resulting constraints  are 
\begin{itemize}
\item $K^0_L \to \mu^+ \mu^-$
\begin{equation}
\frac{1}{M_S^2}\left|
A^{LL}_{\mu\mu ds} +A^{LL}_{\mu\mu sd}
+A^{RR}_{\mu\mu ds} +A^{RR}_{\mu\mu sd}
\right|
\leq
1.8 \times 10^{-4} \mbox{TeV}^{-2},
\end{equation}
\item $K^0_L \to \pi^0  \mu^+ \mu^-$
\begin{eqnarray}
\frac{1}{M_S^4}\Big\{
0.75 \Big[
\big|
A^{LL}_{\mu\mu ds} -A^{LL}_{\mu\mu sd}
\big|^2 
+
\big|
A^{RR}_{\mu\mu ds} -A^{RR}_{\mu\mu sd}
\big|^2 
\Big]
&& \\
-0.48
\mathrm{Re}
\Big[
\big( A^{LL}_{\mu\mu ds} -A^{LL}_{\mu\mu sd} \big)
\big( A^{RR}_{\mu\mu ds} -A^{RR}_{\mu\mu sd} \big)^\ast
\Big]
\Big\}
&\leq&
2.1 \times 10^{-8} \mbox{TeV}^{-4}. \nonumber
\end{eqnarray}
\end{itemize}
Bounds of the order of $100$ TeV are  expected from these
observables.

\subsection{Numerical values}

The above experimental constraints translate into bounds
on the string scale once a model of flavour is chosen.
The coefficients $A_{ijkl}$ are then calculated
in terms of the amplitudes in the original ``flavour'' basis
and the rotation matrices $L,R$. In the lepton sector,
the resulting bounds should be understood as rather naive estimates
since we have not attempted to reproduce the observed lepton
spectrum.

In any case, the consequent bounds are subleading.
Regarding CP violation, we include random order one phases
in the loop corrections (recall that the tree level phases can
be rotated away). This makes the rotation matrices complex
and results in a non--zero Jarlskog invariant.
\begin{table}[h]
\begin{center}
\begin{tabular}{|c|c||c|c|}
\hline
Quark Observables & $M_S  ~({\rm TeV})
$ & (Semi)leptonic Observables &
 $M_S  ~({\rm TeV}) $ \\
\hline
$\Delta m_K$ & $1400$ & $\mu-e$ conversion & 1000 \\
$\Delta m_B$ & $800$ & $\tau\to 3 e$ & 2 \\
$\Delta m_{B_s}$ & $450$ & $\tau\to 3\mu$  & 2 \\
$\Delta m_D$ & $1100$ & $K^0_L\to\mu^+\mu^-$  & 260 \\
$|\epsilon_K|$ & $4\times 10^4$ & $K^0_L\to \pi^0 \mu^+ \mu^-$ & 300
\\ \hline
\end{tabular}
\end{center}
\caption{Bounds on the string scale from flavour violating
  observables.
\label{boundsFCNC:table}}
\end{table}
Using  
Eqs.(\ref{our:parameters}) and (\ref{Lumatrices},\ref{Ldmatrices}), we
obtain the bounds on the string scale shown in
Table.\ref{boundsFCNC:table}. The quark and leptonic
observables are in the left and right columns, respectively.
There are three features worth emphasizing.
First, the bounds are extremely tight. This stems from {\it tree level}
order one flavour violation with order one coupling constants.
Second, the bounds one would naively expect are
indeed realized 
which means that  
no unexpected cancellations are present. 
Finally, although (semi)leptonic observables are less restrictive,
they provide an independent check such that it would be impossible
to circumvent all of the above bounds.

\subsection{Electric dipole moment bounds}

In general, the four--fermion interactions induced by the string instantons
violate CP as well as flavour conservation. CP violation in flavour
changing processes has been considered above, but one also has
constraints from flavour conserving observables such as the 
electric dipole moments.

The typical chirality--flipping 
4--quark  interactions contributing to the atomic and neutron EDMs
(Fig.\ref{LRamplitudenocross.eps}) are of the form
\begin{equation}
\Delta {\cal L}= {\cal Y}_{ijkl}~ \bar q^i_L q^j_R \bar q^k_R q_L^l  
~+~ {\rm h.c.}
\label{edm}
\end{equation}
Here ${\cal Y}_{ijkl}$ is an instanton coupling which contains a
complex phase (as do the Yukawa couplings) due to the presence  of
the antisymmetric background field and Wilson lines
\footnote{We note that this statement  is not generic in string theory. 
Unlike in intersecting brane models, 
only the antisymmetric background field induces physical CP
phases in heterotic string models \cite{Kobayashi:2003gf}.}.
This coupling is suppressed by (roughly) the areas of the quadrangles
spanned by the four vertices in each torus. 

Let us first reiterate how relevant chirality flipping operators
are generated.
Some of the couplings ${\cal Y}_{ijkl}$ reduce to a product of 
the corresponding
Yukawa couplings, i.e. when the ``quadrangles'' are formed by
joining two $\bar q_L q_R H$ triangles through the 
Higgs vertex:
\begin{equation}
\bar q_L^i q_R^j H + \bar q_R^k q_L^l H^* \longrightarrow
\bar q_L^i q_R^j  \bar q_R^k q_L^l \;.
\end{equation}
These ``reducible'' interactions are of no interest to us
since they are real and flavour--diagonal in the basis where
the quark mass matrix is diagonal.
On the other hand, there exist ``irreducible'' contributions
not mediated by the Higgs vertex, which are 
neither flavour--diagonal nor
CP conserving in the physical basis. 
For example, some of these can be 
obtained (schematically)
by combining the Yukawa interactions with the
4--quark chirality conserving operators $\bar q_L^{i} q_L^{i+1}
\bar q_L^{k+1} q_L^{k} $ of Fig.\ref{LLamplitude.eps},
\begin{equation}
\bar q_L^{i+1} q_R^j H + \bar q_R^j q_L^{k+1} H^* +
\bar q_L^{i} q_L^{i+1} \bar q_L^{k+1} q_L^{k}
\longrightarrow
\bar q_L^{i} q_R^{j} \bar q_R^{j} q_L^{k} \;. 
\end{equation}
Since the flavour structure of  $\bar q_L^{i} q_L^{i+1}
\bar q_L^{k+1} q_L^{k}$ is $independent$ of that of the Yukawa
matrices, the resulting interaction is not diagonal in
the physical basis. To get a feeling for the strength
of these interactions, it is instructive to consider
a 2D case, i.e. when the couplings are given by the 
 exponential of the relevant area. Then
\begin{equation} 
{\cal Y}_{ijjk} \sim Y_{kj}/ Y_{ij} \;.
\label{2d}
\end{equation}
This means that ${\cal Y}_{ijkl}$ are not necessarily suppressed
by a product of the quark masses and can lead to significant
constraints on the string scale.
 
Let us consider the down--type  quark sector. The mass eigenstate
basis is given by
\begin{equation}
d_{L}^i = L^{ij} (d_{L})^j_{\rm mass} ~,
\end{equation}
and similarly for the right--handed quarks. In what follows, we will
drop the subscript ``mass''. 
The atomic/neutron EDM measurements constrain most severely
operators of the type $\bar d_L d_R \bar s_R s_L$,
$\bar d_L d_R \bar b_R b_L$, etc. Thus we are interested in
\begin{eqnarray}
\Delta {\cal L} &=& \left( \sum_{i,j,k,l} {\cal Y}_{ijkl}
L^{i1*} R^{j1} R^{k2*} L^{l2}     \right)
\bar d_L d_R \bar s_R s_L ~+~ {\rm h.c.} \nonumber\\
&=& {\rm Im} C_{sd}~
( \bar d i \gamma_5 d ~\bar s s - \bar s i \gamma_5 s ~\bar d d)    
~+~ ({\rm CP-conserving~ piece })
\end{eqnarray}
and analogous terms with $s$$\rightarrow$$b$, etc.  
These CP violating operators contribute to the EDM of the mercury atom
and the neutron \cite{Khriplovich:ga}. 
The matrix element of the second operator in the parentheses
over a nucleon is consistent with zero, while that of the first
operator is known from low energy QCD theorems \cite{shifman}. 
According to the  recent analysis \cite{demir+},
the $d_{\rm Hg}$ bound requires
\begin{equation}
{\rm Im} C_{sd} < 3 \times 10^{-11} ~ {\rm GeV}^{-2} \;.
\label{Csd}
\end{equation}
The bound on ${\rm Im} C_{bd}$ is about an order of magnitude
weaker, while the constraint from $d_n$ is less restrictive.

One can get a rough idea of how restrictive (\ref{Csd}) is by
considering a simple situation in which the rotation matrices $L,R$
are similar to the CKM matrix. Then, an order of magnitude estimate
gives 
\begin{equation}
C_{sd} \sim {1\over 2} {\cal Y}_{1112} ~\sin{\theta_C} \;,
\end{equation}
where ${\cal Y}_{1112}$ generally contains an order one phase and 
can be estimated using (\ref{2d}). 
In practice, when realistic $L,R$ are used and many
terms in $C_{sd}$ are summed, cancellations suppress
${\rm Im} C_{sd}$ by about an order of magnitude and 
the resulting bound is
\begin{equation}
M_S \stackrel{>}{\sim} 10~ {\rm TeV}\;.
\end{equation}
Although this bound is inferior to the FCNC bounds,
it is still important as it implies that some finetuning
is required to obtain a TeV- or 100 GeV-mass Higgs boson 
in non--supersymmetric models (recall
that what matters is the square of the mass).

\section{Supernovae and other constraints}

In the previous sections we derived strong constraints on the string scale
from flavour changing processes and electric dipole moments.
We have utilized a salient feature of the intersecting brane models
that there are 4 fermion flavor changing operators suppressed by $M_S^2$.

In this section, we will employ another rather generic feature of this
class of models, namely, the presence of additional U(1) gauge
bosons and light Dirac neutrinos. These additional U(1) symmetries
are needed to protect a proton from decaying in models with a low
string scale. Further, the intersecting brane constructions naturally lead
to Dirac neutrinos since the neutrinos are localized at brane
intersections and therefore are U(1) charged. The smallness of the
neutrino masses is then explained by an exponential suppression
of the relevant Yukawa couplings and 
a (relatively) large area
of the triangle spanned by the Higgs, left lepton, and right 
neutrino vertices.

We will now use these generic features in order to
obtain additional constraints on the string scale from non--FCNC experiments.
We note that a bound on $M_S$  of order 1 TeV 
from the electroweak $\rho$--parameter has previously been obtained in 
Ref.\cite{Ghilencea:2002da}.
Below we will consider additional phenomenological constraints
based on the presence of extra U(1) gauge bosons and
Dirac neutrinos.  \\

\subsection{Supernova cooling}

The first and potentially strongest constraint comes from
the emission of the right handed neutrinos during supernova collapse,
which affects the rate of supernova cooling \cite{Raffelt:1987yt}.
This places a constraint on the 4 fermion axial vector interaction of the 
form 
\begin{equation}
{\cal L}_R= {4 \pi \over \Lambda^2} ~\bar q \gamma_\mu \gamma_5 q ~
\bar \nu_R \gamma^\mu \nu_R \;. 
\label{SN}
\end{equation}
For 1 type of light Dirac neutrinos, the bound is \cite{Grifols:1997iy}
\begin{equation}
\Lambda \stackrel{>}{\sim} 200~ {\rm TeV}~,
\end{equation}
where we have averaged over degenerate and non-degenerate nucleons.
For 3 types of Dirac neutrinos the bound strengthens by a factor of 
$3^{1/4}$.
A constraint on an analogous vector interaction is weaker ($\sim$
90 TeV) since it leads to a coupling to the nucleon number (rather than
the spin) density which does not fluctuate as much.

The interaction of the type (\ref{SN}) appears in intersecting
brane models. Indeed, since the neutrinos are localized at the
intersections of two branes, they are charged under two U(1)'s.
The corresponding gauge bosons then mediate the contact 
quark--neutrino interaction. The strength of this interaction
depends on two factors: the gauge boson masses and  their
couplings to fermions. These depend on the specifics of the model,
yet the resulting lower bound on the string scale  
can be estimated quite well.

Let us consider the setup of Ref.\cite{Cremades:2002va}.
In this case the so called {\it right} brane connects
the right handed neutrinos and the right handed quarks.
The corresponding ``right'' gauge boson mediates the interaction (\ref{SN})
at the tree level. Actually, to be exact, 
this gauge boson is not a mass eigenstate.
The neutral gauge bosons of the model mix due to the Green--Schwarz
mechanism. In particular, the neutral gauge boson mass matrix is 
given by
\begin{equation}
M^2_{ij}= g_i g_j M_S^2 \sum_{a=1}^3 c^i_a c^j_a \;,
\end{equation}
where $g_i$ is the gauge coupling of U(1)${}_i$, $M_S$ is the string scale,
and $c_a^i$ couples  the U(1)${}_i$ field strength 
to the RR two form field $B_a$ ($a$=1,2,3)
as $c_a^i B_a \wedge {\rm Tr} F^i$ and
 is fixed by the brane wrapping numbers.
The gauge boson mass spectrum in intersecting brane models has 
been studied in Ref.\cite{Ghilencea:2002da}. 
It was found that the lightest mass eigenstate
is lighter than the string scale (up to an order of magnitude), while 
the heaviest eigenstate is heavier than the string scale (up to two
orders of magnitude). It was also found that the light states typically
contain a significant component of the ``right'' or ``leptonic''
gauge bosons.

Further, to compute the contact interaction,
one needs to know the couplings of the U(1) bosons to fermions.
For the model of Ref.\cite{Cremades:2002va}, the anomaly--free hypercharge is written as 
\begin{equation}
q_{_Y}= {1\over 6}q_a -{1\over 2} q_c + {1\over 2} q_d \;,
\end{equation}
where $q_a$ is the charge associated with the U(3) color stack
of branes, $q_c$ -- with the right brane, and $q_d$ -- with the 
leptonic brane. In general, if the charges 
are related by the transformation
\begin{equation}
q_i'= \sum_\alpha U_{i \alpha} q_\alpha \;,
\end{equation}
the corresponding gauge couplings are related by
\begin{equation}
{1\over {g'}_i^{ 2 }}= \sum_\alpha {U_{i \alpha}^2 \over g^2_\alpha} \;.
\end{equation}
Then, the  gauge couplings $g_c$ (``right'')
and $g_d$ (``leptonic'') are found via the relations
\begin{eqnarray} 
{1\over g_Y^2} &=&{1\over 36 g_a^2}+{1\over 4 g_c^2}+{1\over 4 g_d^2}
\nonumber \\
g_a^2 &=& {g_{QCD}^2 \over 6}\;. 
\end{eqnarray}
This means that $g_c$ and $g_d$ cannot be too small, and if one of them
approaches its lower bound, the other one has to be rather large. 
Numerically, we have 
\begin{equation}
{1\over  g_c^2}+{1\over  g_d^2} \sim 40 \;, 
\end{equation}
which implies that $g_c$ and $g_d$ are bounded from below by about 0.15.

\begin{figure}
\caption{\label{supernovatreelevel.eps}
Tree level emission of right--handed neutrinos.}
\begin{center}\includegraphics[%
  scale=0.8]{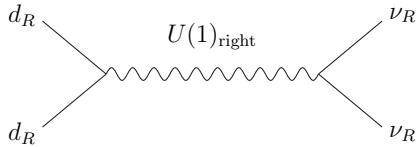}\end{center}
\end{figure}

Now we are ready to estimate the supernova bound on the string scale.
Consider an exchange of a predominantly U(1)$_c$ gauge boson between
$d_R$ and $\nu_R$ (Fig.\ref{supernovatreelevel.eps}). Neglecting the external momenta,
the effective interaction is  
\begin{equation}
{\cal L}_R= {g_c^2 \over 2 M_c^2} ~\bar d \gamma_\mu \gamma_5 d ~
\bar \nu_R \gamma^\mu \nu_R \; + \; {\rm vector ~ piece} \;.  
\end{equation}
To get a conservative bound, set $g_c$ to its minimal value ($\sim$0.15). 
Then
we have $M_c \geq 5$ TeV. As the gauge bosons with a large U(1)$_c$
content are typically lighter than the string scale, this translates into a 
tighter bound on the string scale. On the other hand, the coupling
of the light mass eigenstate to the quarks may be smaller than $g_c$ if there is
a significant  U(1)$_d$ component. Assuming that these two 
effects roughly compensate each 
other, we get 
\begin{equation}
M_S \stackrel{>}{\sim} 5~ {\rm TeV}\;. 
\end{equation}
In more realistic cases when $g_c$ is larger than its minimal allowed value, 
the bound on the string scale lies in the range of tens of TeV 
(e.g. for $g_c \sim g_{_Y}$, $M_S > 10$ TeV). 

\begin{figure}
\caption{\label{supernovaoneloop.eps}
One loop emission of right--handed neutrinos.}
\begin{center}\includegraphics[%
  scale=0.7]{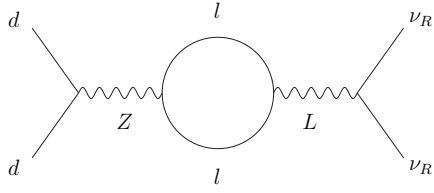}\end{center}
\end{figure}

It is interesting to compare this leading tree level effect to the 
loop--induced quark--right neutrino coupling. It is generated by the 
mixing between the Z boson and the leptonic U(1)$_d$ gauge boson L
via the diagram of Fig.\ref{supernovaoneloop.eps}. This effect is also important 
because
when the coupling $g_c$ is suppressed, $g_d$ is enhanced and the 
1--loop contribution becomes considerable. 

The origin of the Z--L mixing lies in the non--orthogonality of
the Z-- and the leptonic charges,
which is an example of the ``kinetic'' mixing between U(1)'s 
\cite{Holdom:1985ag} (see also recent work \cite{Dienes:1996zr}).
The induced interaction for the d--quarks is of the form
\begin{equation}
{\cal L}_R= {g_Z^A g_d \over  M_Z^2 M_L^2}
\tilde \Pi_{ZL} (q^2=0)
 ~\bar d \gamma_\mu \gamma_5 d ~
\bar \nu_R \gamma^\mu \nu_R \; + \; {\rm vector ~ piece} \;.  
\end{equation}
Here $g_Z^A$ is the d--quark axial Z coupling and  
$\tilde \Pi_{ZL} (q^2=0) $ is the renormalized Z--L  vacuum polarization
at zero momentum transfer (again, L should be understood as a 
predominantly U(1)$_d$ gauge boson).
We note that there is no similar mixing with the photon at zero
momentum transfer by gauge invariance.

To find $\tilde \Pi_{ZL}  $ we use the on--shell renormalization
scheme  for models with multiple U(1)'s  of  Ref.\cite{delAguila:1995rb}.
The renormalization conditions are imposed on the inverse gauge
boson propagator in the Feynman gauge
\begin{eqnarray}
D_{\mu \nu}^{-1} (q^2)&=&i g_{\mu \nu} \left(
\begin{array}{cc} 

q^2-M_Z^2 + \tilde \Pi_{ZZ}(q^2) & \tilde \Pi_{ZL}(q^2)\\
             \tilde \Pi_{ZL}(q^2) & q^2-M_L^2 + \tilde \Pi_{LL}(q^2)
\end{array} 
\right)  \nonumber\\
&=& i g_{\mu \nu} \left(  
\begin{array}{cc}   D_{ZZ}^{-1} & D_{ZL}^{-1} \\
                    D_{ZL}^{-1} & D_{LL}^{-1} 
\end{array}\right) \;, 
\end{eqnarray}
where we have omitted the $q_\mu q_\nu$ part of the vacuum polarization
which would lead to the amplitude suppressed by the fermion masses.
The six on--shell conditions read
\begin{eqnarray}
&& D_{ZZ}^{-1} \Bigl\vert_{q^2=M_Z^2}=D_{ZL}^{-1} \Bigl\vert_{q^2=M_Z^2}
   =0 \;\;,\;\; {\partial  D_{ZZ}^{-1} \over \partial q^2} \Biggl\vert_{q^2=M_Z^2}=1 \;\;, \nonumber \\
&& D_{LL}^{-1} \Bigl\vert_{q^2=M_L^2}=D_{ZL}^{-1} \Bigl\vert_{q^2=M_L^2}
   =0 \;\;,\;\; {\partial  D_{LL}^{-1} \over \partial q^2} \Biggl\vert_{q^2=M_L^2}=1 \;\;.
\end{eqnarray}
They can be satisfied by choosing 3 appropriate ``wave function'' and
3 ``mass'' counterterms. The conditions for the off--diagonal propagator
decouple form the others:
\begin{eqnarray}
\label{ren}
&& \tilde \Pi_{ZL}(M_Z^2)=\tilde \Pi_{ZL}(M_L^2)=0 \;, \\
&& \tilde \Pi_{ZL}(q^2)= q^2 \Pi_{ZL}(q^2) +Z_{ZL} q^2 + \delta M^2_{ZL}
   = {\rm finite} \;,
\end{eqnarray}
where $Z_{ZL}$ is the mixed ``wave function'' and $\delta M^2_{ZL}$
is the mixed ``mass'' counterterms (which do not enter into the diagonal conditions). As we are interested in $\tilde \Pi_{ZL}(0)$, we only need to
compute $\delta M^2_{ZL}$. 

The quantity $\Pi_{ZL}(q^2)$ is given by the bubble diagram with 
the Z and L external legs. It is actually identical to the one 
in QED  (apart from the gauge couplings) due to the vectorial nature
of the L boson. For one fermion in the loop and 
neglecting the lepton masses, we have 
\begin{equation} 
\Pi_{ZL}(q^2)= {g_Z^V g_d \over 6 \pi^2} \biggl[
{1\over \epsilon} + c -{1\over 2} \ln \left( {-q^2 \over \mu^2} \right)
\biggr] \;,
\end{equation}
where $g_Z^V$ is the vector Z coupling, $1/ \epsilon + c$ is 
the UV part of the diagram, and $\mu$ is the renormalization scale. 
Solving (\ref{ren}) for $\delta M^2_{ZL}$ and summing over all leptons,
we get
\begin{equation}
\delta M^2_{ZL}= \left(  \sum_\ell (g_Z^V)_\ell \right) g_d~
{1\over 12 \pi^2} {M_Z^2 M_L^2 \over M_L^2 -M_Z^2} \ln {M_Z^2 \over M_L^2}\;.
\end{equation}
Then the resulting quark--neutrino interaction can be written as
\begin{equation}
{\cal L}_R \simeq {g^2 g_d^2 \tan^2 \theta_W  \over  16 \pi^2 M_L^2}
\ln {M_Z^2 \over M_L^2}
 ~\bar d \gamma_\mu \gamma_5 d ~
\bar \nu_R \gamma^\mu \nu_R \; + \; {\rm vector ~ piece} \;.  
\end{equation}
For $g_d \sim 0.2$  this places a bound on $M_L$ of about 1 TeV,
which according to the above arguments we can translate into 
\begin{eqnarray}
M_S \stackrel{>}{\sim} 1~ {\rm TeV}\;. 
\end{eqnarray}
This result can be expected on general grounds since this bound is roughly
a loop factor down compared to the tree level one. 

Our conclusion here is that the supernova cooling constraint places
 a bound on the string scale of about 5 TeV at the tree level. 
If for some (special) reason the leading effect is suppressed,
the loop induced amplitude constrains $M_S$ to be above 1 TeV.

\subsection{LEP constraints on contact interactions}

The OPAL measurements of the angular distributions of 
the $e^+ e^- \rightarrow \ell^+ \ell^- $ processes  \cite{Abbiendi:1998ea}
  constrain
the non--SM contact interactions of the type
\begin{equation}
{\cal L}_R= {4 \pi \over \Lambda^2} ~\bar e \gamma_\mu  e ~
\bar \ell \gamma^\mu \ell \;, 
\end{equation}
with 
\begin{equation}
\Lambda > 9.3 ~{\rm TeV}
\end{equation}
at 95\% CL. Other types of contact interactions, i.e. axial,
chirality specific, etc., are constrained slightly weaker. 
An exchange of the L gauge boson produces the vector--vector 
contact interaction above with $4\pi / \Lambda^2 = g_d^2/ M_L^2$,
which  for $g_d \sim 0.2$ translates into 
\begin{equation}
M_S \stackrel{>}{\sim} 0.5~ {\rm TeV}\;.
\end{equation}
In most of the parameter space ($g_d > g_d^{\rm min}$),
the bound is in the 1 TeV range. This is comparable to the 1$\sigma$
bound from the $\rho$ parameter obtained in Ref.\cite{Ghilencea:2002da}.

\subsection{$\rho$ parameter.}

This constraint has been studied in Ref.\cite{Ghilencea:2002da}.
The basic idea here is that the physical Z boson has
an admixture of ``heavy'' U(1) gauge bosons with 
St\"uckelberg masses. On the other hand, the W boson
does not get a similar contribution. This modifies
the Standard Model relation between the Z and W masses,
i.e. affects the $\rho$--parameter
\begin{equation}
\rho = {M_W^2 \over M_Z^2 \cos^2 \theta_W }\;.
\end{equation}
This leads to a (1$\sigma$) bound on the string scale 
in the range of 1 TeV,
\begin{equation}
M_S \stackrel{>}{\sim} 1~ {\rm TeV}\;.
\end{equation}
There are further constraints similar in strength which rely
on the presence of extra U(1) gauge bosons (see Ref.\cite{Leike:1998wr}
for a comprehensive discussion).

\section{Conclusions}

In this work, we have studied the  (related) issues of flavour and constraints
on the string scale in intersecting brane models.
Flavour is known to be a problematic point in these constructions.
In particular, the Yukawa factorization property implies that
the first two fermion families are massless and there is no 
quark mixing. 
In this paper, we pointed out that this is only true at tree level.
The ever--present higher dimensional operators generated by
string instantons contribute to the Yukawa couplings through
one loop threshold corrections. As a result, the Yukawa factorizability
is lost and realistic flavour structures are possible even in the
simplest (non-supersymmetric) models. The lightness of
the first two generations is then explained by the loop suppression.
We note that this mechanism also fixes the relevant compactification radii
to be around the string length.

Further, we addressed the issue of whether the string scale in intersecting
brane models can be in the TeV range, as required in non--supersymmetric
models. To answer this question, we employ generic features of this class of 
models such as the mechanism of fermion family replication, the presence
of extra gauge bosons and Dirac neutrinos, etc. to obtain 
phenomenological constraints on the string scale. 
We found that the strongest bounds are due to the flavour
changing neutral currents which appear at $tree$ level. 
To suppress them sufficiently, 
the string scale has  to be higher than $10^4$ TeV.
This bound has been derived using the flavour structure alluded to above,
in which case the dominant FCNC contributions are induced by string instantons.
In principle, if some compactification radii were relatively large, the dominant
contribution would be provided by the gauge KK mode exchange
and the bound on the string scale would relax to ${\cal O}(100)$ TeV
\cite{Abel:2003fk}. However,
it would be difficult (if possible at all) to obtain a realistic flavour
pattern in this case.

Bounds on the string scale from other experiments are weaker.
Nonobservation of the EDMs constrains CP--violating
flavour--conserving operators, resulting in the bound $M_S > 10$ TeV
(this bound is also sensitive to the model of flavour).
Emission of right--handed neutrinos during supernova SN1987A collapse
places a bound $M_S > 5-10$ TeV independently of the flavour model.
Collider and $\rho$--parameter bounds are inferior, $M_S > 1$ TeV.

The above bounds are sufficiently strong to rule out 
a 1 TeV string scale.
This allows us to conclude that non--supersymmetric intersecting brane models
face a severe fine--tuning problem and supersymmetry is needed
to address the hierarchy problem. We note that supersymmetry is also
favoured by stability considerations. 

It is not clear at the moment whether fully
realistic flavour structures can arise in \textit{supersymmetric} 
configurations.
First of all, the intersection angles are more constrained. Also,
due to the SUSY non--renormalization theorem, the 
threshold corrections to the Yukawa couplings from 4 point operators  
would behave as $m_{\rm SUSY}^2/M_S^2$ after SUSY breaking,
which is too small to generate the light quark masses. 
Nevertheless, we point out that there is another source of flavour
structures in SUSY models. The flavour pattern of the A--terms is
different  from that of the Yukawa matrices and the corresponding
SUSY vertex corrections can generate non--zero masses for the light
generations. Whether or not this mechanism produces a fully 
realistic spectrum remains to be seen.

\section*{Acknowledgements}

It is a pleasure to thank I. Navarro and B. Schofield for useful
discussions. This work was funded by PPARC and by Opportunity Grant 
PPA/T/S/ 1998/00833.

\section*{Appendices}

\section*{A: The classical contribution to the amplitude}

The equations of motion, asymptotic behaviour at intersections and monodromy conditions 
are  sufficient to determine the classical 
instanton $X_{cl}$ and its corresponding action. 
The action is expressed in terms of 
the hypergeometric functions
\begin{equation}
\label{eq:fs}
\begin{array}{l}
F_{1}=e^{-i \pi(\vartheta_{2}+\vartheta_{3})}x_{2}^{-1+\vartheta_{1}+\vartheta_{2}}B(\vartheta_{1},\vartheta_{2}){}_{2}F_{1}(\vartheta_{1},1-\vartheta_{3},\vartheta_{1}+\vartheta_{2};x),
  \\
F_{2}=e^{-i \pi(-1+\vartheta_{3})}(1-x_{2})^{-1+\vartheta_{2}+\vartheta_{3}}B(\vartheta_{2},\vartheta_{3}){}_{2}F_{1}(\vartheta_{3},1-\vartheta_{1},\vartheta_{2}+\vartheta_{3};1-x).
\end{array}
\end{equation}
We also define 
\begin{eqnarray}
\alpha &=& -\frac{\sin(\pi(\vartheta_{1}+\vartheta_{2}))}{\sin(\pi\vartheta_{1})} ,\nonumber \\
\beta  &=& -\frac{\sin(\pi(\vartheta_{2}+\vartheta_{3}))}{\sin(\pi\vartheta_{3})} ,\nonumber\\
\gamma &=& \frac{\Gamma(1-\vartheta_{2})\Gamma(1-\vartheta_{4})}{\Gamma(\vartheta_{1})
\Gamma(\vartheta_{3})}, \nonumber \\
\gamma' &=& \frac{\Gamma(\vartheta_{2})\Gamma(\vartheta_{4})}{\Gamma(1-\vartheta_{1})
\Gamma(1-\vartheta_{3})},
\end{eqnarray}
and 
\begin{equation}
\label{eq:tau}
\tau(x)=\left|\frac{F_{2}}{F_{1}}\right|.
\end{equation}
The contribution to the action from a single sub-torus is found to be 
\begin{equation}
\label{4ptaction}
S^{T_{2}}_{cl}(\tau,v_{21},v_{32})=
\frac{\sin(\pi\vartheta_{2})}{4\pi\alpha'}\left(\frac{((v_{21}\tau-v_{32})^{2}+
\gamma\gamma'(v_{21}(\beta+\tau)+v_{32}(1+\alpha\tau))^{2})}
{(\beta+2\tau+\alpha\tau^{2})}\right),
\end{equation}
where $v_{21}$ and $v_{32}$ are the lengths of sides $12$ and $23$ in that 
particular subtorus. 
The complete expression for the classical action is just the sum of these
contributions, one from each torus subfactor. Hence
\begin{equation}
\label{sumscl}
S_{cl}=\sum_{m=1}^{3}S^{T^{m}_{2}}_{cl}(\tau^{m},v^{m}_{21},v^{m}_{32}).
\end{equation} 

The final amplitude (see below) involves an integration 
over $x$ of the final expression which includes the factor $exp(-S_{cl}(x))$,  
and so is dominated by saddle point contributions given by 
(if the polygon is convex in all subtori)
\begin{equation} \frac{\partial S_{cl}}{\partial x} = 0.
\end{equation}
If the angles (and hence $\tau^m$) are the same in every torus,  
we may easily get the saddle point using the condition  
\begin{equation}
 \frac{\partial S_{cl}}{\partial \tau} = 0
\end{equation}
instead, which has a simple functional form. In the case that the ratios of 
sides $v^m_{12}/v^m_{23}$ are degenerate as well, the action reduces to the 
sum of the projected areas as in the 3 point case if the polygon is convex. 
However, this is a very special situation, and when it is not the case
we find a source for new flavour structure.

\section*{B: The complete amplitude and $s$- and $t$-channel Higgs exchange }

The above expressions together with the quantum piece  
may be used to determine any 4 point amplitude. 
Here, for the generation of a one-loop threshold contribution to the Yukawas,  
we will present the interactions that 
mimic Higgs exchange, with two left-handed and two right-handed fermions. 
First, we collect the quantum contributions to the amplitude. 
These are \begin{eqnarray}
ghosts\times\langle e^{-\phi/2}(0)e^{-\phi/2}(x)e^{-\phi/2}(1)e^{-\phi/2}(x_{\infty})\rangle & = & x_{\infty}^{\frac{1}{2}}x^{-\frac{1}{4}}(1-x)^{-\frac{1}{4}},\nonumber \\
\nonumber \\\langle e^{-ip_{1}.X}e^{-ip_{2}.X}e^{-ip_{3}.X}e^{-ip_{4}.X}\rangle & = & x^{2\alpha'p_{1}.p_{2}}(1-x)^{2\alpha'p_{2}.p_{3}},\nonumber \\
\langle e^{iq_{1}.H}e^{iq_{2}.H}e^{iq_{3}.H}e^{iq_{4}.H}\rangle_{cmp} & = & \prod_{m}^{3}x_{\infty}^{\vartheta_{4}^{m}(1-\vartheta_{4}^{m})-\frac{1}{4}}x^{\vartheta_{1}^{m}\vartheta_{2}^{m}-\frac{1}{2}(\vartheta_{1}^{m}+\vartheta_{2}^{m})+\frac{1}{4}}\nonumber \\
 & \times  & (1-x)^{\vartheta_{2}^{m}\vartheta_{3}^{m}-\frac{1}{2}(\vartheta_{2}^{m}+\vartheta_{3}^{m})+\frac{1}{4}} ~,\end{eqnarray}
 where the last piece is for the three compactified tori factors only.
The final piece comes from the uncompactified part of the fermions
and is chirality dependent. 
We are interested in the operator $(\overline{q}_{L}^{(3)}q_{R}^{(2)})(\overline{e}_{R}^{(1)}e_{L}^{(4)})$
for which the uncompactified part of the fermions
 have charges (denoted $\tilde{q}_{i}$)
\begin{eqnarray}
\widetilde{q_{1}} & = & -\tilde{q}_{4}=\pm(\frac{1}{2},\frac{1}{2}),\nonumber \\
\widetilde{q_{2}} & = & -\tilde{q}_{3}=\pm(\frac{1}{2},-\frac{1}{2}).\end{eqnarray}
 Identifying the two $\pm$ possibilities with the Weyl spinor indices
$\alpha$ of the fermions $u_{\alpha}$, we see that
in $\overline{u}_{\dot{\alpha}}^{(3)}\overline{u}_{\dot{\beta}}^{(2)}u_{\gamma}^{(1)}u_{\delta}^{(4)}$
we have opposite $\dot{\alpha}\dot{\beta}$ and $\gamma\delta$
indices which (writing as $\varepsilon_{\dot{\alpha}\dot{\beta}}\varepsilon_{\gamma\delta}$)
just contract the $\overline{q}_{L}q_{R}$ and $\overline{e}_{L}e_{R}$
fermions. 
We then find
\begin{eqnarray}
\langle e^{-iq_{1}.H}e^{-iq_{2}.H}e^{-iq_{3}.H}e^{-iq_{4}.H}\rangle_{non-cmp} & = & x_{\infty}^{\tilde{q}_{4}.(\tilde{q}_{1}+\tilde{q}_{2}+\tilde{q}_{3})}x^{\tilde{q}_{1}.\tilde{q}_{2}}(1-x)^{\tilde{q}_{2}.\tilde{q}_{3}} \\
 & = & x_{\infty}^{-\frac{1}{2}}x^{\tilde{q}_{1}.\tilde{q}_{2}}(1-x)^{\tilde{q}_{2}.\tilde{q}_{3}}=x_{\infty}^{-\frac{1}{2}}(1-x)^{-\frac{1}{2}} .
\nonumber\end{eqnarray}
 The additional half-integer power in the $t$-channel is required 
 when we extract the Higgs pole in the $x\rightarrow1$ limit.

Upon adding in the contribution from the
bosonic twist fields as in 
Refs.\cite{Cvetic:2003ch}-\cite{Abel:2003yx}, 
we find
that dependence
on $\vartheta_{i}^{m}$
cancels between the bosonic twist fields and the spin-twist fields.
The final expression for the amplitude is (in four component
notation)
\begin{equation}
\begin{array}{lll}
A(1,2,3,4) & = & -g_{s}\alpha'\,\int_{0}^{1}dx\,\, x^{-1-\alpha's}(1-x)^{-1-\alpha't}\frac{(1-x)^{-\frac{1}{2}}}{\prod_{m}^{3}|J^{m}|^{1/2}}\\
 &  & \,\,\,\,\times\left[(\overline{u}^{(3)}u^{(2)})(\overline{u}^{(1)}u^{(4)})\right]\sum e^{-S_{cl}(x)}~.
\end{array}\end{equation}
These amplitudes lead to the expected $s-$ and $t-$ channel Higgs exchanges.
The Higgs field appears at the $SU(2)$-$U(1)$ brane intersection
and so this process is the equivalent of the $t$-channel Higgs exchange
in the field theory limit.  The exchange appears as a double instanton as shown in figure \ref{qqll} with the Higgs state appearing as a pole.
The $s$-channel exchange corresponds to the 
opposite ordering of vertices (so $x\rightarrow 1-x $).
Since the instanton suppression
goes as $e^{-Area/2\pi\alpha'}$ we expect to find the product of
two Yukawa couplings. The amplitude should go as \[
\frac{Y_{u}Y_{e}}{t-M_{h}^{2}}\]
or the $s$ channel equivalent. We can verify this behaviour as follows. 

It is easy to show that when the diagram has an intersection as in
Fig.\ref{qqll2}, the action is monotonically decreasing and we
may approximate the integral by taking the limit $x\rightarrow1$.
Assuming that $1-\vartheta_{2}-\vartheta_{3}>0$, the relevant limits
are\begin{eqnarray}
Lim_{x\rightarrow1}(\tau) & = & -\beta ~,\nonumber \\
Lim_{x\rightarrow1}(J) & = & (1-x)^{(-1+\vartheta_{2}+\vartheta_{3})}\,\frac{1}{\gamma}\,\,\eta(\vartheta_{2},\vartheta_{3})\eta(1-\vartheta_{1},1-\vartheta_{4})~,\end{eqnarray}
where \begin{equation}
\eta(\vartheta_{i},\vartheta_{j})=\left(\frac{\Gamma(\vartheta_{i})\Gamma(\vartheta_{j})\Gamma(1-\vartheta_{i}-\vartheta_{j})}{\Gamma(1-\vartheta_{i})\Gamma(1-\vartheta_{j})\Gamma(\vartheta_{i}+\vartheta_{j})}\right)^{\frac{1}{2}}~.
\end{equation}
 The normalization of the amplitudes and Yukawas can be obtained in
this limit as in Ref.\cite{Cvetic:2003ch}. We take the limit where the 4
point function with no intersection turns into  the 3 point function. 
The
normalization factor for the general 4 point function is \begin{equation}
2\pi\prod_{m=1}^{3}\sqrt{\frac{4\pi}{\gamma_{m}}\,\frac{\eta(1-\vartheta_{2}^{m},1-\vartheta_{3}^{m})}{\eta(\vartheta_{1}^{m},\vartheta_{4}^{m})}}\end{equation}
and the Yukawas take the form found in Ref.\cite{Cvetic:2003ch},
\begin{equation}
Y_{23}(A_{m})=16\pi^{\frac{5}{2}}\prod_{m}^{3}\eta(1-\vartheta_{2}^{m},1-\vartheta_{3}^{m})\,\sum_{m}e^{-A_{m}/2\pi\alpha'} ~, 
\end{equation}
where $A_{m}$ is the projected area of the triangles in the $m$'th 
2-torus. Once we add the intersection, the interior $\vartheta_{1},\,\vartheta_{4}$
angles become exterior and should be replaced by 
$1-\vartheta_{1},\,1-\vartheta_{4}$,
respectively. The constraint on the interior angles is now $\vartheta_{1}+\vartheta_{4}=\vartheta_{2}+\vartheta_{3}$
because of the intersection. Looking at $Lim_{x\rightarrow1}(J)$
we see that this can be taken into account by adding an extra $\sqrt{\eta(1-\vartheta_{1},1-\vartheta_{4})/\eta(\vartheta_{1},\vartheta_{4})}=\eta(1-\vartheta_{1},1-\vartheta_{4})$
factor to the 4 point amplitude. We then find\begin{equation}
S_{4}=\alpha'\, Y_{23}(0)\, Y_{14}(0)\, e^{-S_{cl}(1)}\,\int_{0}^{1}dx\,(1-x)^{-\alpha't-\sum\frac{1}{2}(\vartheta_{2}^{m}+\vartheta_{3}^{m})}.\end{equation}
The contribution to the
classical action from each sub-torus becomes\begin{equation}
S_{cl}(1)=\frac{1}{2\pi\alpha'}\left(\frac{\sin\pi\vartheta_{1}\sin\pi\vartheta_{4}}{\sin(\pi\vartheta_{1}+\pi\vartheta_{4})}\frac{v_{14}^{2}}{2}+\frac{\sin\pi\vartheta_{2}\sin\pi\vartheta_{3}}{\sin(\pi\vartheta_{2}+\pi\vartheta_{3})}\frac{v_{23}^{2}}{2}\right).\end{equation}
We see that the result is just the sum of the area/$2\pi\alpha'$
of the two triangles. Finally, the pole term now arises from the $x$
integral,
\begin{equation}
\alpha'\int_{0}^{1}dx\,(1-x)^{-\alpha't-\sum\frac{1}{2}(\vartheta_{2}^{m}+\vartheta_{3}^{m})}=\frac{1}{t-M_{h}^{2}}~,
\end{equation}
where (recalling that $0<\vartheta_{2}^{m}+\vartheta_{3}^{m}<1$)
we recognize the mass of a scalar Higgs state in the spectrum at the
intersection,
\begin{equation}
\alpha'M_{h}^{2}=1-\frac{1}{2}\sum_{m}(\vartheta_{2}^{m}+\vartheta_{3}^{m}).\end{equation}
 The opposite ordering of operators leads to the $s$- channel exchange
in the $x\rightarrow0$ limit. The above discussion was carried out
for intersecting D6-branes, but it is straightforward to translate
it to other set-ups.

\end{document}